\documentclass[a4paper,11pt]{article}

\usepackage{jcappub} 

\usepackage[T1]{fontenc}
\usepackage[utf8]{inputenc}
\usepackage{bbding} 
\usepackage{soul}
\usepackage{cancel}

\usepackage{amsmath, amsfonts, amssymb}
\usepackage{graphicx,float,color}
\usepackage{hyperref}
\hypersetup{colorlinks = true, linkcolor = blue, urlcolor  = blue, citecolor = blue, anchorcolor = blue}
\usepackage{multirow}
\usepackage[numbers,sort&compress]{natbib}
\usepackage{braket}
\usepackage{epic}
\usepackage{eepic}
\usepackage{epsfig}
\usepackage{latexsym}
\usepackage{color}
\usepackage{url}  
\usepackage{caption}
\usepackage[caption=false]{subfig}
\usepackage{float}
\usepackage{bm}
\usepackage{geometry}
\usepackage{slashed}
\usepackage[detect-all]{siunitx}

\usepackage{physics}
\usepackage{amsmath} 
\usepackage{amsfonts}
\usepackage{color}
\usepackage{url}
\usepackage{graphicx,float,color}
\usepackage{verbatim}

\usepackage{tablefootnote}
\usepackage{aas_macros}

\newcommand{\crd}{}

\usepackage{etoolbox}

\title{\boldmath Probing Boson Clouds with Supermassive Black Hole Binaries}

\author[a,b]{Ximeng Li}
\author[a,c]{Jing Ren}
\author[a,b]{Xi-Li Zhang}

\affiliation[a]{Institute of High Energy Physics, Chinese Academy of Sciences, Beijing 100049, China}
\affiliation[b]{School of Physics Sciences, University of Chinese Academy of Sciences, Beijing 100039, China}
\affiliation[c]{Center for High Energy Physics, Peking University, Beijing 100871, China}

\emailAdd{liximeng@ihep.ac.cn}
\emailAdd{renjing@ihep.ac.cn}
\emailAdd{zhangxili@ihep.ac.cn}

\abstract{
Rotating black holes can generate boson clouds via superradiance when the boson's Compton wavelength is comparable to the black hole's size. In binary systems, these clouds can produce distinctive observational imprints. Recent studies accounting for nonlinearities induced by orbital backreaction suggest that if the binary forms at a large separation, resonant transitions can significantly deplete the cloud, minimizing later observational consequences except for very specific orbital inclinations. In this paper, we extend this framework to supermassive black hole binaries (SMBHBs), considering the influence of their astrophysical evolutionary histories. We find that, before entering the gravitational wave (GW) radiation stage, the additional energy loss channels can accelerate orbital evolution. This acceleration makes hyperfine resonant transitions inefficient, allowing a sufficient portion of the cloud to remain for later direct observations. We further discuss the ionization effects and cloud depletion occurring at this stage. Based on these theoretical insights, we explore how multi-messenger observations for SMBHBs can be utilized to detect the ionization effects of boson clouds by examining changes in the orbital period decay rate via electromagnetic measurements and variations in GW strain over a wide frequency band.
Our findings reveal a complex dependence on the binary's total mass, mass ratio, and boson mass, emphasizing the significant role of astrophysical evolution histories in detecting boson clouds within binaries.

}

\arxivnumber{2505.02866}

\begin{document}
\maketitle
\flushbottom

\section{Introduction}

Ultralight bosons are well-motivated in particle physics, with examples such as the QCD axion, proposed to resolve the strong CP problem, and axion-like particles, which can be produced through the compactification of extra dimensions in string theory. Black hole (BH) superradiance offers an intriguing method to probe these bosons, even if they couple only gravitationally. In the presence of massive bosons, BH superradiance is turned into an instability, allowing a boson cloud to naturally form around a rotating black hole, resulting in what is known as a ``gravitational atom''. When the Compton wavelength of the boson is comparable to the BH size--specifically, when the gravitational fine structure constant $\alpha=GM\mu\sim\mathcal{O}(0.1)$ {\crd in natural units $c=\hbar=1$}, where $M$ is the BH mass and $\mu$ is the boson mass--the boson can grow efficiently by extracting the mass and angular momentum of BH. Thus, astrophysical observations of stellar mass BHs with $M\sim \mathcal{O}(1)M_{\odot}$ to supermassive BHs with $M\sim \mathcal{O}(10^{10})M_{\odot}$ offer a unique window to probe ultralight bosons with masses in the range of $\mu\sim 10^{-21}-10^{-11}$\,eV (see Ref.~\cite{Brito:2015oca} for a review). 

The saturated boson cloud around a single BH can be probed indirectly via BH spindown or directly through monochromatic gravitational waves (GWs) emitted by the cloud~\cite{Arvanitaki:2009fg, Arvanitaki:2010sy}. If the gravitational atom is inside a binary, the binary companion can perturb the cloud's {\crd initial state}, yielding resonant transitions to other bound states~\cite{Baumann:2018vus,Zhang:2018kib,Baumann:2019ztm,Berti:2019wnn}, ionization to unbound states~\cite{Baumann:2021fkf, Takahashi:2021yhy, Baumann:2022pkl}, dynamical friction and accretion onto the companion~\cite{Ferreira:2017pth, Zhang:2019eid,Baumann:2021fkf,Tomaselli:2023ysb}. These phenomena can affect the binary orbit, leaving distinct imprints on the inspiral evolution of the BH binary. As the binary evolves from its formation, if it starts at a sufficiently large separation, resonant transition to a decaying bound state may occur, potentially deplete the cloud efficiently. Some earlier studies suggested that too little cloud might remain after the resonant transition~\cite{Berti:2019wnn}, making it insufficient for direct detection through ionization at later stages.  
However, a recent study~\cite{Tomaselli:2024bdd} explored the nonlinear evolution of the cloud-binary system for a generic orbit and found that the cloud could survive in very specific orbital configurations, offering the possibility of direct detection later on. Despite these findings, most discussions assume that the BH binaries exist in a vacuum, where GW radiation is the only energy loss channel. In reality, the astrophysical evolution history of BH binaries could play a crucial role in the evolution of the cloud-binary system, as discussed in the case of stellar mass BH binaries in Ref.~\cite{Guo:2024iye}.


In this paper, we apply the analysis framework developed in Ref.~\cite{Tomaselli:2024bdd} to supermassive black hole binaries (SMBHBs), considering the potential influence of their astrophysical evolution histories. 
We focus on SMBHBs formed from gas-poor galaxy mergers, with total masses ranging from $M\sim 10^{6}-10^{10}M_{\odot}$ and located at low redshifts{--given their relevance to electromagnetic and gravitational wave observations}. 
We further assume that the heavier BH hosts a boson cloud, while the lighter one, characterized by a mass ratio of $q\sim 0.01-0.1$, does not.\footnote{This is because, for $\alpha \lesssim 0.3$, the corresponding fine structure constant for the lighter BH, i.e. $q\alpha$, would be quite small, resulting in an extremely long instability timescale for its cloud.} The mass ratio is chosen in this range to ensure the validity of the perturbative treatment within the analysis framework~\cite{Tomaselli:2024bdd} and to guarantee that SMBHBs can indeed form through galaxy mergers~\cite{2002MNRAS.331..935Y}.
Moreover, for SMBHBs located at low redshifts, the effects of gas can generally be ignored~\cite{Chen:2020qlp}, 
{ which in turn prevents significant SMBH mass increase from accretion during the cloud evolution and guarantees the applicability of the theoretical framework in Ref.~\cite{Tomaselli:2024bdd}.}

The formation and evolution of a SMBHB in a gas-poor galaxy merger can be divided into several stages~\cite{2002MNRAS.331..935Y}. Initially, after the merger of the two galaxies, the BHs move closer together from $\mathcal{O}(\rm kpc)$ to $\mathcal{O}(\rm pc)$ distances due to dynamical friction. Eventually, the binary forms a bound state, during which dynamical friction decreases and three-body interactions with stars increasingly contribute to energy loss. 
{\crd The binary then progresses into the hard binary stage, where interactions with stars can provide an important channel for energy loss, though other mechanisms such as interactions with gas may also contribute.}
Finally, once the SMBHB separation becomes sufficiently small, GW radiation dominates, transitioning the system into the GW radiation stage. 
Resonant transitions, such as hyperfine transitions with very small energy differences, are triggered by orbital motion with low angular frequencies at large separations. Depending on the value of $\alpha$, these transitions may occur at sufficiently large separations before the binary enters the GW radiation stage. This suggests that conclusions about the remaining cloud may differ significantly from those in Ref.~\cite{Tomaselli:2024bdd}, when accounting for additional energy loss channels during the binary's evolution. This will then influence the detection sensitivity to boson clouds through observations of SMBHBs. 

This paper is structured as follows. 
In Sec.~\ref{sec:theory}, we focus on the theoretical aspects of scalar boson cloud effects on SMBHB inspirals. Specifically, 
{ adopting the simplest model of a free scalar boson without self-interaction,}
we examine how the astrophysical evolution history of SMBHBs influences resonant transitions, particularly hyperfine transitions, and affects the amount of the remaining boson cloud in Sec.~\ref{sec:resonance}. We then discuss ionization effects and cloud depletion during ionization in greater detail in Sec.~\ref{sec:ionization}. In Sec.~\ref{sec:observation}, we explore potential constraints on the ionization effects of boson clouds through multi-messenger observations of SMBHBs. 
We summarize our findings in Sec.~\ref{sec:summary}.
{\crd Throughout the paper, we adopt natural units, $c=\hbar=1$, unless stated otherwise.} 

\section{Impact of boson clouds on SMBHB inspiral}
\label{sec:theory}

Scalar bosons can extract energy and angular momentum from a rotating BH through superradiant scattering, provided the scalar's frequency $\omega$ is less than the angular velocity of the BH's event horizon $\Omega_+=\chi/(2r_+)$, i.e. $\omega<m \Omega_+$. Here, $m$ is the azimuthal quantum number, $\chi$ is the BH dimensionless spin and $r_+=GM(1+\sqrt{1-\chi^2})$ is the horizon radius. If the scalar boson has a nonzero mass $\mu$, this superradiant scattering can lead to an instability, resulting in the formation of a boson cloud around the BH. This process is most efficient when the Compton wavelength of the bosons, $1/\mu$, is comparable to the gravitational radius of the BH, $GM$. In the far-field region, the Klein-Gordon equation simplifies to a Schrödinger equation, leading to the BH-boson cloud system being referred to as a ``gravitational atom.'' The physics of this system is primarily governed by the gravitational fine structure constant $\alpha \equiv GM\mu$. 

Similar to a hydrogen atom, 
the bound states of a gravitational atom can be denoted as $|a\rangle=|n\ell m\rangle$, where $n$ is the  principal quantum number and $\ell$ denotes the orbital angular momentum. However, unlike a hydrogen atom, the scalar boson cloud has complex eigenvalues for the bound states, i.e. $\omega_{a}=E_{a}+i \Gamma_{a}$. In the limit $\alpha\ll 1$, the real and imaginary parts can be approximated as \cite{Detweiler:1980uk, Baumann:2018vus, Baumann:2019eav}
\begin{eqnarray} \label{eq: energy_level}
 E_{a}&\approx &\mu \left[ 1-\frac{\alpha ^2}{2n^2}-\frac{1}{n^4}\left( \frac{1}{8}+\frac{3n-2\ell -1}{\ell +1/2} \right) \alpha ^4+\frac{2m}{n^3\ell \left( \ell +1/2 \right) \left( \ell +1 \right)}\chi \alpha ^5+\mathcal{O} \left( \alpha ^6 \right) \right]\nonumber\\
 \Gamma_{a}&\approx & 2r_+C_{n\ell} g_{\ell m}(\chi, \alpha, \omega) (m\Omega_+-\mu)\alpha^{4\ell+5}+\mathcal{O}(\alpha^{4\ell+7})\,,
\end{eqnarray}
where the numerical coefficients $C_{n\ell}$ and $g_{\ell m}$ can be found in Ref.~\cite{Baumann:2019eav}. 
The binding energy is defined as $\epsilon_{a}=E_{a}-\mu<0$. 
It is helpful to define the Bohr radius for gravitational atoms as $r_c=GM/\alpha^2$.
The normalized wave function for the eigenstate is then expressed as
\begin{equation}\label{eq:wavef}
    \phi_{a}(\mathbf{r})=\mu^{3/2}R_{n\ell}(r)Y_{\ell m}(\theta, \psi)
    = \frac{\alpha^3}{(GM)^{3/2}}\bar{R}_{n\ell}(x)Y_{\ell m}(\theta, \psi)\,,
\end{equation}
where $R_{nl}$ is the radial function and $Y_{\ell m}$ is the spherical harmonic. The second expression makes explicit the scaling behavior of the radial function in terms of the dimensionless radius $x\equiv r/r_c$, i.e.
\begin{equation}  \label{eq: Rnl}
    \bar{R}_{n\ell}\left( x\right) =\sqrt{\left( \frac{2}{n} \right) ^3\frac{\left( n-\ell -1 \right) !}{2n\left( n+1 \right) !}}\mathrm{e}^{-\frac{x}{n}}\left( \frac{2x}{n} \right) ^{\ell}L_{n-\ell -1}^{2\ell +1}\left( \frac{2x}{n} \right)\,.
\end{equation}
At leading order in the $\alpha$ expansion, the energy density of the boson cloud is approximately proportional to $\mu^2 |\phi_{a}|^2$. The prefactor is determined by the condition $\int \rho_{a}(\mathbf{r})\mathrm{d}^3\mathbf{r} =M_c$, where $M_c$ denotes the cloud mass. Thus, the energy density can be approximated as 
\begin{align}\label{eq:rhoa}
    \rho_{a}(\mathbf{r})=\alpha^6{\crd \frac{1}{G^3M^2}}\bar\rho_{a}(x,\theta,\phi),\quad \bar\rho_{a}(x,\theta,\phi)=\bar{M}_c |\bar{R}_{n\ell}(x)Y_{\ell m}(\theta,\phi)|^2
\end{align}
where $\bar{M}_c=M_c/M$, and $\bar\rho_{a}$ depends only on $x$ and the angular variables.

The gravitational atom also has continuous unbound states, denoted as $|K\rangle=|k;\ell m\rangle$, with $k$ representing the wavenumber. The energy of these states is expressed as  $E_K=\sqrt{\mu^2+k^2}$, and the wave function is given by
\begin{eqnarray}
\phi_{K}(\mathbf{r})=R_{k;\ell}(r)Y_{\ell m}(\theta, \psi)=   
\frac{\alpha^2}{GM}\bar{R}_{\bar{k}; \ell}(x)Y_{\ell m}(\theta, \psi)\,.
\end{eqnarray}
Here, $\bar{R}_{\bar{k}; \ell}(x)$ denotes the dimensionless radial function in terms of the dimensionless radius $x$ and the dimensionless wavenumber  $\bar{k}\equiv GM k/\alpha^2$, with
\begin{equation}\label{eq:Rkl}
\bar{R}_{\bar{k}; \ell}(x)=\frac{ 2
 e^{\frac{\pi}{2 \bar{k}}}\left|\Gamma\left(\ell+1+i\bar{k}^{-1}\right)\right|}{(2 \ell+1)!}(2 \bar{k} x)^{\ell} e^{-i \bar{k} x}{ }_1 F_1\left(\ell+1+i\bar{k}^{-1} ; 2 \ell+2 ; 2 i \bar{k} x\right)\,.
\end{equation}
Here, ${ }_1 F_1(a;b;z)$ is the Kummer confluent hypergeometric function. Since continuous states do not exhibit exponential decay at spatial infinity, normalization occurs only after integrating over $k$.      
In $\bar{k}\ll 1$ and $\bar{k}x\ll 1$ limit, the radial function can be approximated as 
\begin{eqnarray}\label{eq:Rklsmallk}
\bar{R}_{\bar{k};\ell}(x)\approx 
\sqrt{\frac{4\pi \bar{k}}{x}}J_{2\ell+1}(2\sqrt{2x})\,,
\end{eqnarray}
where $J_{\nu }(z)$ is the Bessel function of the first kind.

{\crd It is reasonable to assume that  the fastest-growing superradiant mode, $|n\ell m\rangle=|211\rangle$, serves as the initial state of the boson cloud. For a free scalar boson without self-interaction, this mode has an instability timescale given by}
\begin{eqnarray}\label{eq:tauinst}
\tau_{\rm inst}=\frac{1}{|\Gamma_{211}|}\approx 48GM \alpha^{-9}\left(\chi m-2\mu r_+\right)\,,
\end{eqnarray}
The boson cloud can experience significant growth only if $\tau_{\rm inst}$ is considerably less than the typical time scale for the BH evolution,
which then set a lower bound of $\alpha\gtrsim \alpha_{\rm inst}$. Considering near-extremal BHs with  $\chi\approx 1$, the explicit values of $\alpha_{\rm inst}$ for the SMBH mass range of interest are listed in Tab.~\ref{tab:alpha}.\footnote{{ Note that it takes about $100\tau_{\rm inst}$ for the  boson cloud to saturate from a small initial quantum fluctuation. However, due to the $\alpha^{-9}$ dependence of $\tau_{\rm inst}$ in Eq.~(\ref{eq:tauinst}), altering $\tau_{\rm inst}$ to $100\tau_{\rm inst}$ would only modify $\alpha_{\rm inst}$ by 60\%.  Given that our later chosen benchmark values for $\alpha$ are considerably larger than $\alpha_{\rm inst}$, such an $\mathcal{O}(1)$ correction has minimal impact on our subsequent discussion.}}
{ The cloud growth reaches saturation when the black hole's spin decreases to the critical value $\chi_s$, at which point the condition for superradiance amplification is no longer met, i.e. when $\omega=m\chi_s/(2r_+)$. Given that $\omega\approx \mu$, this leads to $\chi_s m=2 \mu r_+$, 
i.e. $\chi_s\approx (4/m)\alpha$ for $\alpha\ll 1$.
The saturated mass of the cloud is around $\bar{M}_c\approx \alpha$~\cite{Brito:2015oca}.}  The cloud then gradually loses energy and angular momentum through GW emission. The typical timescale for this process is \cite{Yoshino:2013ofa, Brito:2014wla}
\begin{eqnarray}\label{eq:tauGW}
\tau_{\rm GW}\approx 81GM \alpha^{-15}\,.
\end{eqnarray}
Given the typical evolution time scale, this then determines a critical value of $\alpha_{\rm GW}$, as shown in Tab.~\ref{tab:alpha}, above which the decay of the cloud via GW emission must be considered. The time evolution of the cloud mass is given by~\cite{Brito:2014wla} 
\begin{eqnarray}  \label{eq:Mc_GW_decay}
\bar{M}_c(t)\approx \frac{\alpha}{1+t/\tau_{\rm GW}}\,.
\end{eqnarray} 

When a gravitational atom has a BH companion, gravitational perturbations from the companion can induce transitions from the {\crd initial state} to either bound or unbound states. 
Assuming that the lighter BH {\crd is located} at $\{r(t),\theta(t),\psi(t)\}$ relative to the heavier one, the leading order gravitational perturbation from the companion is described by the potential~\cite{2008gady.book.....B, Baumann:2019ztm}
\begin{equation} \label{eq: tidal_V}
V\left(t, \mathbf{r}' \right) =-\sum_{\ell=0}^{\infty}\sum_{m=-{\ell}}^{{\ell}}\frac{4\pi q \alpha}{2\ell +1}{\crd Y_{\ell m}\left( \theta ,\psi \right) Y_{\ell m}^{*}\left( \theta',\psi' \right) }F\left(r, r'
\right),
\end{equation}
where $Y_{\ell m}$ {\crd are the} spherical harmonics and 
\begin{eqnarray}  \label{eq: F(r)}
\begin{aligned}
F\left(r, r' \right) \equiv 
\frac{\alpha^2}{GM}\bar{F}\left(x, x' \right)= \frac{\alpha^2}{GM}\begin{cases}
	\dfrac{x'^{\ell}}{x^{\ell +1}}\Theta \left( x-x' \right) +\dfrac{x^{\ell}}{x'^{\ell +1}}\Theta \left( x'-x \right) , \ell \neq 1\,,\\
	\left( \dfrac{x}{x'^{2}}-\dfrac{x'}{x^2} \right) \Theta \left( x'-x \right) , \ell =1\,,\\
\end{cases}
\end{aligned}
\end{eqnarray}
where $\Theta$ is the Heaviside function. This potential can induce transitions from the {\crd initial state} to other bound states, which are further classified into Bohr ($\Delta n\neq 0$), {\crd fine ($\Delta n=0, \Delta \ell\neq 0$)}, {\crd hyperfine ($\Delta n=0, \Delta \ell=0, \Delta m\neq 0$)} transitions~\cite{Baumann:2019ztm}. It can also cause transitions to unbound states, a process referred to as ionization~\cite{Baumann:2021fkf}.

Resonant transitions can be triggered only when the { binary's orbital angular frequency, $\Omega$,}
matches the energy difference between the two transitioning states. According to Eq.~(\ref{eq: energy_level}), the hyperfine transition, which has the smallest energy difference, occurs first during the early inspiral stage, followed by the fine transition. After {\crd the} extended GW radiation stage, Bohr transition and ionization become relevant in the late inspiral phase. 
In the following subsections, we will first discuss resonant transitions for SMBHBs, taking into account their astrophysical evolutions. We will then focus on ionization effects. For both subsections, we will highlight the potential differences from previous studies. 


\subsection{Transition to bound states}
\label{sec:resonance}

The transition from the {\crd initial state} $|
a\rangle$ to another bound state $|
b\rangle$ is governed by the matrix element of the potential term in Eq.~(\ref{eq: tidal_V}). Angular momentum conservation dictates that the transition probability is nonzero only when a set of selection rules are satisfied, i.e. $m=m_b-m_a$, $\ell=\ell_a+\ell_b+2p$, where 
$p=0,\pm1, \pm2, \cdots$ and $-\left|\ell_b-\ell_a\right|\le \ell\le \left|\ell_b+\ell_a\right|$. Expanding the spherical harmonic $Y_{\ell m}(\theta,\psi)$ in terms of the form $Y_{\ell g}(\pi/2,0)$ and substituting in the wave function in Eq.~(\ref{eq:wavef}), the matrix element can be expressed as~\cite{Tomaselli:2024bdd}
\begin{equation}  \label{Fourier_coef}
\left< a \right|V(t) \left| b \right> 
=\sum_{g=-\ell}^{\ell}{\eta ^{\left( g \right)}(t)\, \mathrm{e}^{\mathrm{i}g\int\mathrm{d}t \Omega}}\,,
\end{equation}
where {\crd $\Omega$ is the binary's orbital angular frequency.} Multiple $g\in\mathbb{Z}$ terms can contribute due to nonzero eccentricity $e$ and inclination $\iota$, and 
\begin{equation} \label{eq: eta}
    \eta^{(g)}(t) = 4\pi q\alpha \sum_{g=-\ell}^{\ell} {\frac{d_{m,g}^{\ell}(\iota(t))}{2\ell +1}  Y_{\ell g}\left(\frac{\pi}{2},0\right)I_r(t)I_{\Omega}}\,,
\end{equation}
where $d_{m,g}^{\ell}$ is a Wigner small $d$-matrix. 
The radial integration factor is given by 
\begin{eqnarray}\label{eq:Ir}
I_r(t)&=&\int_0^{\infty}{\mu^3 F\left(r(t), r^{\prime} \right) R_{n_a\ell _a}\left( r^{\prime} \right) R_{n_b\ell _b}\left( r^{\prime} \right) {r^{\prime}}^2\mathrm{d}r^{\prime}}\nonumber\\
&=&\frac{\alpha^2}{GM} \int_0^{\infty}\bar{R}_{n_a\ell_a}(x') \bar{R}_{n_b \ell_b}(x')\bar{F}(x,x')  x'^2 dx'\,,
\end{eqnarray}
with $\bar{R}_{nl}(x)$ and $\bar{F}(x,x')$ give in Eq.~(\ref{eq: Rnl}) and (\ref{eq: F(r)}), respectively.
The angular variable integration factor $I_{\Omega}$ is nonzero when the selection rules are satisfied.


Without backreaction, the two-state transition can be described by a {\crd Schr\"{o}dinger} equation.
A resonant transition occurs when {\crd $g\Omega=g\Omega_0\equiv \Delta E$}, where $\Delta E=E_b-E_a$ is the energy difference between the two states. Near the resonant frequency $\Omega_0$, the angular frequency evolves approximately linearly with time, 
\begin{eqnarray}\label{eq:rateG}
\bar\Omega\approx \bar\Omega_0+\mathcal{G}\bar{t}\,, 
\end{eqnarray}
where $\bar\Omega=\Omega GM$, $\bar{t}=t/(GM)$ are dimensionless parameters, and $\mathcal{G}$ describes the orbital evolution rate. If the orbital decay is solely determined by GW radiation, the rate is given by
\begin{eqnarray}\label{eq:Ggr}
    \mathcal{G}= \frac{96}{5} q\,\bar{\Omega}_0^{11/3}f(e)\,,
\end{eqnarray}
where $f(e)\equiv (1-e^2)^{-7/2}(1+73/24e^2+37/96e^4)$. For more general cases involving additional energy loss channels, it is useful to relate $\mathcal{G}$ to the evolution timescale for the binary, defined as $t_{\rm evol}=|a/(da/dt)|$, where $a$ is the semi-major axis of the orbit. Using the Keplerian relation $\bar\Omega^2\bar{a}^3=1$, where $\bar{a}=a/(GM)$ {(for circular orbits, we define $\bar{r}=r/(GM)$)}, 
we can find
\begin{eqnarray}\label{eq:Gtevl}
    \mathcal{G}=\frac{3}{2}\frac{\bar\Omega_0}{\bar{t}_{\rm evol}}\,,
\end{eqnarray}
where $\bar{t}_{\rm evol}=t_{\rm evol}/(GM)$. Around the resonant frequency, the Schr\"{o}dinger equation can be written as~\cite{Tomaselli:2024bdd}
\begin{eqnarray}  \label{eq: dim0_two_states}
    \frac{\mathrm{d}}{\mathrm{d}\bar{t}}\left( \begin{array}{c}
	c_a\\
	c_b\\
\end{array} \right) =-\mathrm{i}\left( \begin{matrix}
	g\mathcal{G} \bar{t}/2&		\bar{\eta}^{(g)}\\
	\bar{\eta}^{(g)} &		-g\mathcal{G} \bar{t}/2-\mathrm{i}\bar{\Gamma}_b\\
\end{matrix} \right) \left( \begin{array}{c}
	c_a\\
	c_b\\
\end{array} \right) ,
\end{eqnarray}
where $\bar{\eta}^{(g)}\equiv GM\eta^{(g)}$   
and $\bar{\Gamma}_b=GM \Gamma_b$ are the dimensionless counterparts. Here, the {\crd initial state} $\left|a\right>$ has vanishing decay width due to the saturation condition, while the excited state $\left|b\right>$ has a nonzero decay rate $\Gamma_b>0$ as it is reabsorbed by the central BH. As the diagonal terms increase, a Landau-Zener two-state transition could be triggered if the following condition is satisfied  \cite{landau2013quantum, doi:10.1021/jp040627u}
\begin{equation}\label{eq:LZZ}
    \mathcal{Z  }\equiv  \frac{\left(\bar{\eta}^{\left( g \right)} \right) ^2}{\left| g \right|\mathcal{G}}\gg \frac{1}{2\pi}.
\end{equation}
In this regime, a small sweeping rate $\left|g\right|\mathcal{G}$ allows sufficient time for the transition to occur, resulting in nearly all of the cloud transferring to the $\left|b\right>$ state. In the opposite limit, i.e. $2\pi \mathcal{Z}\ll 1$, the transition becomes non-adiabatic and is significantly suppressed.

It turns out that these results can be strongly affected by the backreaction of the boson cloud on orbital evolution. Specifically, when accounting for backreaction, the temporal evolution of the orbital frequency, eccentricity, and inclination angle, as governed by energy and angular momentum conservation, is described by~\cite{Tomaselli:2024bdd}
\begin{subequations}\label{eq: float_dynamics}
\begin{align}
\frac{\mathrm{d}\bar{\Omega}}{\mathrm{d}\bar{t}}&=\mathcal{G} -\mathcal{B}\frac{\mathrm{d}\left| c_b \right|^2}{\mathrm{d}\bar{t}}, \label{eq: dOmega_dt}
\\
\frac{\mathrm{d}e}{\mathrm{d}\bar{t}}&=\frac{\sqrt{1-e^2}}{3e\bar{\Omega}}\left( -\sqrt{1-e^2}\frac{\mathrm{d}\bar{\Omega}}{\mathrm{d}\bar{t}}-\frac{\Delta m}{g}\mathcal{B}\frac{\mathrm{d}\left| c_b \right|^2}{\mathrm{d}t}\cos \iota 
+ \Delta X_e \right), \label{eq: de_dt} 
\\
\frac{\mathrm{d}\iota}{\mathrm{d}\bar{t}}&=-\frac{1}{3\bar{\Omega}\sqrt{1-e^2}}\frac{\Delta m}{g}\mathcal{B}\frac{\mathrm{d}\left| c_b \right|^2}{\mathrm{d}\bar{t}}\sin \iota , \label{eq: diota_dt}
\end{align}
\end{subequations} 
where $\Delta X_e$ denotes the contribution from a generic dissipation mechanism. For GW radiation, $\Delta X_e=\mathcal{G} h\left( e \right)$, with $h(e)=(1-e^2)^{-2}(1+7/8 e^2)$. The strength of the backreaction is characterized by the dimensionless parameter 
\begin{eqnarray}\label{eq:brB}
\mathcal{B}\equiv -\frac{3g}{q\alpha}\bar{M}_c \bar{\Omega}_0^{4/3}\,.
\end{eqnarray}
When $\mathcal{B}>0$, the backreaction can expand the parameter space for an adiabatic transition. In particular, when $2\pi \mathcal{Z}\ll 1$, the transition can become adiabatic again if $2\pi\mathcal{Z} \mathcal{B}\sqrt{|g|/\mathcal{G}}>1$, and it remains non-adiabatic only if $2\pi\mathcal{Z} \mathcal{B}\sqrt{|g|/\mathcal{G}}<1$~\cite{Tomaselli:2024bdd}. 

When the condition for adiabatic transition is met, the binary system enters a floating orbit stage where $\Omega$ no longer evolves, i.e. $d|c_b|^2/d\bar{t}=\mathcal{G}/\mathcal{B}$ from Eq.~(\ref{eq: dOmega_dt}). For $\Delta |c_b|^2\sim 1$, the timescale for the floating orbit is approximately
\begin{eqnarray} \label{eq: t_float}
\Delta \bar{t}_{\mathrm{float}}
\approx \frac{\mathcal{B}}{\mathcal{G}}
=\frac{9(-g)}{2}\frac{\bar{M}_c}{\alpha}q^{-1}\bar{\Omega}_0^{1/3}\bar{t}_{\rm evol} \,.
\end{eqnarray} 
In the final expression, we establish a connection between the floating time and the evolution timescale for the binary using Eq.~(\ref{eq:Gtevl}).
Given that $\bar\Omega_0$ is very small for hyperfine transitions, for $q\sim 0.01-0.1$, it generally follows that $\Delta \bar{t}_{\mathrm{float}}<\bar{t}_{\rm evol}$. 
During the floating time, the eccentricity and inclination angle continue to evolve according to Eqs.~(\ref{eq: de_dt}) and (\ref{eq: diota_dt}), with the rate $d|c_b|^2/dt$ determined by the floating condition. This evolution leads to distinctive features in these two quantities, depending on specific mechanisms of energy and angular momentum loss.

A floating orbit can break when the time variations of various parameters become significant. Among all, the breaking due to the non-negligible decay width of the excited state  $\left|b\right>$ in Eq.~(\ref{eq: dim0_two_states}) is most relevant~\cite{Tomaselli:2024bdd}. If its decay timescale is sufficiently shorter than $\Delta t_{\mathrm{float}}$, the transition will end when the remaining population in the {\crd initial state} becomes 
\begin{equation}\label{eq:Gammabreaking}
  \left| c_{a,1} \right|^2\approx \frac{\bar\Gamma_b}{2\mathcal{Z}\mathcal{B}}\,.
\end{equation}
If the floating orbit breaks early enough, a significant portion of the cloud may remain. The cloud mass available for subsequent evolution is given by $\bar{M}_{c,1}\approx \alpha |c_{a,1}|^2$.


Now, let's examine the consequences of hyperfine transitions for SMBHBs, considering the evolution history effects when the binaries have large separations before entering the GW radiation stage.  For the initial state $|a\rangle=\left|211\right>$, a hyperfine transition can occur to $|b\rangle=\left|210\right>$ or $\left|21\textrm{-}1\right>$, with $\Delta m =m_b-m_a=-1$ or $-2$, respectively, where the former is the fastest decaying mode. 
According to the selection rule, for both cases, $\ell$ must be an even number satisfying $\ell>|\Delta m|$, and $0\le\ell\le2$, meaning the potential receives contributions only from the $\ell=2$ mode. Furthermore, the sum over $g$ is dominated by the mode with $g=-\ell=-2$, where the Wigner $d$-matrix element $d_{m,g}^{\ell}(\iota)$ is proportional to $\sin^{\Delta m+2}(\iota/2)\cos^{-\Delta m+2}(\iota/2)$. The backreaction parameter $\mathcal{B}$ in Eq.~(\ref{eq:brB}) is then positive, enabling the possibility of a floating orbit. 

By substituting the small energy splitting from Eq.~\eqref{eq: energy_level} and using the Keplerian relation, the orbital angular frequency and semi-major axis for the hyperfine transition are given by  
\begin{equation}  \label{eq: initial_omega}
\bar{\Omega} _0=\frac{1}{24} \alpha ^6\chi_s|\Delta m|,\quad
\bar{a}_0=24^{2/3}\alpha^{-4}\chi_s^{-2/3}|\Delta m|^{-2/3}\,.
\end{equation}
{With $\bar{a}_0 \equiv a_0/GM$ }To derive the expansion parameter $\eta^{(g)}$ for the hyperfine transition in Eq.~(\ref{eq: eta}), we first need to calculate the radial integration factor $I_r$ in Eq.~(\ref{eq:Ir}). Since the wavefunction of the initial state peaks around $x\sim \mathcal{O}(1)$, while the hyperfine transition occurs at a very large radius $x_0=\bar{a}_0\alpha^2\propto \alpha^{-2}\chi_s^{-2/3}\gg 1$, the integration is dominated by the first term of $\bar{F}(x_0,x')$ in Eq.~(\ref{eq: F(r)}). This yields $I_r\propto \alpha^2 x_0^{-3}\propto \alpha^{8}\chi_s^2$, and thus the expansion parameter in Eq.~(\ref{eq: eta}) can be simplified as 
\begin{eqnarray}\label{eq:eta211}
\bar\eta^{(g)} \approx 
c_{\Delta m}\, q \alpha^{9}\chi_s^2|\Delta m|^2 {\crd H(\iota)},
\end{eqnarray}
where $c_{-1}\approx0.024$ and $c_{-2}\approx 0.014$. The dependence on the inclination angle is encoded in 
\begin{eqnarray}
    {\crd H(\iota)}\equiv \sin^{2+\Delta m} (\iota /2) \cos^{2-\Delta m}( \iota /2 )\,,
\end{eqnarray}
which arises from the rotation associated with the Wigner $d$-matrix. The other two quantities relevant for the adiabaticity conditions in Eqs.~(\ref{eq:LZZ}) and (\ref{eq:brB}) are then given by
\begin{eqnarray}\label{eq:ZB211}
    \mathcal{Z} \approx 
    \frac{1}{2}c_{\Delta m}^2 q^2\alpha ^{18} \chi_s^4 |\Delta m|^4\mathcal{G}^{-1} {\crd H(\iota)},\quad
    \mathcal{B} \approx 0.087\, q^{-1}\alpha ^{7}\chi_s^{4/3}|\Delta m|^{4/3} \bar{M_c}\,. 
\end{eqnarray}
The strong dependence of $\mathcal{Z}$ on  the inclination angle $\iota$ means that the efficiency of adiabatic transitions in the floating orbit is sensitive to $\iota$. 
For later discussion, it is convenient to rewrite the two key quantities associated with the adiabatic transition conditions as 
\begin{eqnarray}
2\pi\mathcal{Z}\equiv \mathcal{C}_1 {\crd H(\iota)},\quad
2\pi\mathcal{Z} \mathcal{B}\sqrt{\frac{|g|}{\mathcal{G}}}\equiv \mathcal{C}_2 {\crd H(\iota)}\,,
\end{eqnarray}
where the two constants are defined as
\begin{eqnarray}\label{eq:C1C2}
\mathcal{C}_1\approx 
\pi c_{\Delta m}^2q^2\alpha ^{18} \chi_s^4 |\Delta m|^4\mathcal{G}^{-1},\;
\mathcal{C}_2\approx 
0.39 c_{\Delta m}^2\bar{M}_c \,q\,\alpha ^{25} \chi_s^{16/3} |\Delta m|^{16/3}\mathcal{G}^{-3/2}\,.
\end{eqnarray}
Since $|{\crd H(\iota)}|\leq 1$, the non-adiabatic conditions are satisfied for any orbit when the constants $\mathcal{C}_1, \mathcal{C}_2\ll1$. Otherwise, the non-adiabatic conditions may impose non-trivial constraints on $\iota$ with ${\crd H(\iota)}\lesssim 1/\mathcal{C}_1, 1/\mathcal{C}_2$.


\begin{table}[h]
\begin{center}
\begin{tabular}{c|ccccc|cc}
\hline\hline
& & & & & & &
\\[-3mm]
$M/M_{\odot}$ &  $\alpha_{\rm inst}$   & $\alpha_{\rm bound}$   & 
$\alpha_{\rm NA}$   & $\alpha_{\rm GW}$  & $\alpha_{\rm gr}$ 
& $a_{\rm gr}$\,(pc) & $t_{\rm peak}$\,(Gyr) \\
& & & & & & &
\\[-3.5mm]
\hline
& & & & & & &
\\[-3mm]
$10^6$ & 0.03 & 0.05 & 0.08 & 0.12 &  0.17& $5.5\times 10^{-4}$ &  $0.9$
\\
& & & & & & &
\\[-3.5mm]
\hline
& & & & & & &
\\[-3mm]
$10^7$ & 0.03 & 0.05 & 0.09 & 0.12  &  0.19& $4\times 10^{-3}$ &  2.6
\\
& & & & & & &
\\[-3.5mm]
\hline
& & & & & & &
\\[-3mm]
$10^8$ & 0.04 & 0.05 & 0.10 & 0.14 & 0.21& $2.2\times 10^{-2}$ &  4.9
\\
& & & & & & &
\\[-3.5mm]
\hline
& & & & & & &
\\[-3mm]
$10^9$ & 0.05 & 0.05 & 0.12 & 0.17 & 0.23 & $1.4\times 10^{-1}$ &  5.2
\\
& & & & & & &
\\[-3.5mm]
\hline
& & & & & & &
\\[-3mm]
$10^{10}$ & 0.06 & 0.05 &  0.13 & 0.18 & 0.26&  $8.9\times 10^{-1}$ & 5.5
\\[-3mm]
& & & & & & &
\\
\hline\hline
\end{tabular}
\caption{Summary of various critical values of $\alpha$ for the {\crd initial state} $|a\rangle=|211\rangle$ within the SMBH mass range of interest. Here, $\alpha_{\rm inst}$ and $\alpha_{\rm GW}$ are determined by equating the instability time scale $\tau_{\rm inst}$ in Eq.~(\ref{eq:tauinst}) and GW emission time $\tau_{\rm GW}$ in Eq.~(\ref{eq:tauGW}) to the typical timescale $t_{\rm peak}$ (last column) for the evolution of SMBHBs, respectively.  $\alpha_{\rm bound}$ in Eq.~(\ref{eq:alphabd}) and $\alpha_{\rm gr}$ in Eq.~(\ref{eq:alphagr}) are linked to the boundaries that the binary become gravitationally bound and GW radiation begins to dominate the evolution of SMBHBs when the hyperfine transition occurs. $\alpha_{\rm NA}$ defines the parameter space for non-adiabatic hyperfine transitions, as specified in Eq.~(\ref{eq:alphana}). The last two columns show the peak values of $a_{\rm gr}$ and $t_{\rm peak}$ from statistical distributions of the SMBHBs' evolution studied in Ref.~\cite{Chen:2020qlp}\protect\footnotemark.  
Note that, due to the strong $\alpha$ dependence of the relevant conditions, the critical values of $\alpha$ shown here are quite insensitive to the specific value of $\Delta m$ or $q\sim 0.01-0.1$. } 
\label{tab:alpha}
\end{center}
\end{table}
\footnotetext{For $M\sim 10^6-10^9 M_{\odot}$, the values of $a_{\rm gr}$ and $t_{\rm peak}$ are given for binaries with $q=1$~\cite{Chen:2020qlp}, based on the galaxy distribution model from Ref.~\cite{2014MNRAS.444.3340W}.
For $0.01\lesssim q\lesssim 0.1$, $a_{\rm gr}$ will be smaller by less than an order of magnitude, resulting in a slightly larger $\alpha_{\rm gr}$. The values of $a_{\rm gr}$ and $t_{\rm peak}$ for $M=10^{10}M_{\odot}$ are provided through private communication with the authors of Ref.~\cite{Chen:2020qlp}.
}
We are now ready to explore the implications of the above discussions to more realistic SMBHBs. A binary becomes gravitationally bound  when $a<a_{\rm bound}\approx GM/\sigma_c^2$, { where we adopt $\sigma_c\approx 200\,\textrm{km\,s}^{-1}$ (i.e. $\sigma_c\approx6.7\times 10^{-4}$ in natural units) as the benchmark} velocity dispersion of {\crd the} merged galaxy core {\crd }~\cite{Chen:2020qlp}.\footnote{{ The velocity dispersion $\sigma_c$ can be estimated via the empirical relation~\cite{Gallazzi:2006zt}: $\log(\sigma_c/\text{km\,s}^{-1}) = 0.286\log(M_*/M_\odot)-0.895$, where $M_*$ denotes the bulge mass of the galaxy merger remnant. For SMBH masses $M\sim 10^{6}-10^{10}\,M_\odot$, the BH–bulge mass relation from Ref.~\cite{Chen:2020qlp} gives $M_*\sim 10^{10}-10^{12}\,M_\odot$, yielding $\sigma_c\sim 100-300\,\text{km\,s}^{-1}$. Given the $\sigma_c^{1/2}$ dependence of $\alpha_{\rm bound}$, the order-one difference from the benchmark value $\sigma_c\sim 200\,\text{km\,s}^{-1}$ has a negligible impact.}} 
A critical value of $\alpha$ is then defined through $\bar{a}_0=\bar{a}_{\rm bound}$, i.e. 
{\crd
\begin{eqnarray}\label{eq:alphabd}
    \alpha_{\rm bound}\approx 1.7\, \sigma_c^{1/2} \chi_s^{-1/6}|\Delta m|^{-1/6}\,,
\end{eqnarray}}
which is independent of $M$ and $q$.
When $\alpha\gtrsim \alpha_{\rm bound}$, the hyperfine transition occurs after the binary enters the non-hard binary stage with $a_0<a_{\rm bound}$. The radius at which binaries enter the GW radiation stages, 
i.e. $a_{\rm gr}$, does not scale with $GM$ due to complex hardening processes, such as three-body interactions with stars. For a given $a_{\rm gr}$, we can determine a critical value $\alpha_{\rm gr}$ from $\bar{a}_0=\bar{a}_{\rm gr}$, i.e.
{\crd
\begin{eqnarray}\label{eq:alphagr}
    \alpha_{\rm gr}\approx 1.7\, \bar{a}_{\rm gr}^{-1/4} \chi_s^{-1/6}|\Delta m|^{-1/6}\,.
\end{eqnarray}}
When $\alpha\lesssim \alpha_{\rm gr}$, environmental effects on SMBHB evolution must be considered, and the rate $\mathcal{G}$ in Eq.~(\ref{eq:rateG}) should include other contributions to energy loss. From Eq.~(\ref{eq:Gtevl}), this  corresponds to a shorter evolution timescale $t_{\rm evol}$ compared to the case where only GW radiation is considered. 
If the hyperfine transition occurs after the SMBHB becomes bound but before entering the GW radiation stage, i.e. $
\alpha_{\rm bound}\lesssim 
\alpha\lesssim \alpha_{\rm gr}$, we can establish a conservative lower bound on $\mathcal{G}$ using $t_{\rm evol}<t_{\rm peak}$ from Eq.~(\ref{eq:Gtevl}), i.e.
\begin{eqnarray}\label{eq:Gtpeak}
 \mathcal{G}\gtrsim \frac{3}{2}\frac{\bar\Omega_0}{\bar{t}_{\rm peak}}\,,
\end{eqnarray}
where $t_{\rm peak}$ denotes the maximum evolution time during this period~\cite{Chen:2020qlp}. If the hyperfine transition occurs even before the binary becomes bound, i.e., $\alpha\lesssim 
\alpha_{\rm bound}$, we instead use $t_{\rm evol}<t_{\rm uni}$, where $t_{\rm uni}\approx 1.4\times 10^{10}\,$yr is the age of the universe. The lower bound on $\mathcal{G}$ is then given by, 
\begin{eqnarray}\label{eq:Gtuni}
 \mathcal{G}\gtrsim \frac{3}{2}\frac{\bar\Omega_0}{\bar{t}_{\rm uni}}\,.
\end{eqnarray}
Since the two quantities $\mathcal{C}_1$ and $\mathcal{C}_2$ in Eq.~(\ref{eq:C1C2})
are both inversely proportional to $\mathcal{G}$, a lower bound on $\mathcal{G}$ provides a conservative estimate for meeting the non-adiabatic conditions.

The evolution of SMBHBs in galaxies with realistic property distributions has been explored in Ref.~\cite{Chen:2020qlp}, which presented statistical properties of various quantities associated with SMBHB evolution after the binaries become bound. 
Using the peak values of $a_{\rm gr}$ from these distributions,  
we determine the critical value of $\alpha_{\rm gr}$  in Eq.~(\ref{eq:alphagr}) for hyperfine transition, as shown in Tab.~\ref{tab:alpha}. 
For $\alpha_{\rm bound}\lesssim 
\alpha\lesssim \alpha_{\rm gr}$, we find the first non-adiabatic condition $\mathcal{C}_1\ll 1$ always satisfied from Eqs.~(\ref{eq:C1C2}) and (\ref{eq:Gtpeak}). For the second non-adiabatic condition, assuming $\bar{M}_c\sim \alpha$ for $\mathcal{B}$ as a conservative estimate, we find that $\mathcal{C}_2\lesssim 1$ only if $\alpha\lesssim \alpha_{\rm NA}$, where $\alpha_{\rm NA}$ sets the critical value for the validity of the non-adiabatic evolution. From Eqs.~(\ref{eq:C1C2}) and (\ref{eq:Gtpeak}), this critical value is given by 
{\crd 
\begin{eqnarray}\label{eq:alphana}
\alpha_{\rm NA}\approx 0.82\, c^{-1/8}_{\Delta m}\bar{M}_c^{-1/16}q^{-1/16}\chi_s^{-23/96}|\Delta m|^{-23/96}\bar{t}_{\rm peak}^{-3/32}\,.
\end{eqnarray}}
With the peak values of $t_{\rm peak}$ from the SMBHBs' distributions, we find the explicit values of $\alpha_{\rm NA}$, as shown in Tab.~\ref{tab:alpha}.
For $\alpha\lesssim \alpha_{\rm bound}$, with the very loose condition $t_{\rm evol}<t_{\rm uni}$, we find both $\mathcal{C}_1$ and $\mathcal{C}_2$ are much smaller than one from Eqs.~(\ref{eq:C1C2}) and (\ref{eq:Gtuni}). Thus, the environmental effects on SMBHBs before they enter the GW radiation stage are sufficient to make the hyperfine transition non-adiabatic for $\alpha_{\rm inst}\lesssim\alpha\lesssim 
\alpha_{\rm NA}$.

For $\alpha\gtrsim\alpha_{\rm NA}$, the second condition imposes non-trivial constraints on $\iota$. Specifically, the function ${\crd H(\iota)}$ in Eq.~(\ref{eq:C1C2}) can introduces a strong suppression in two regimes:
for $\iota\to 0$, ${\crd H(\iota)}\approx (\iota/2)^{4+2\Delta m}$, and for $\iota\to 180^{\circ}$, ${\crd H(\iota)}\approx ((180^{\circ}-\iota)/2)^{4-2\Delta m}$. Since the hyperfine transition occurs for both $\Delta m=-1$ and $-2$, efficient suppression is achieved only when the orbit is nearly counter-rotating, i.e. $\iota\to 180^{\circ}$.  
For $\alpha\gtrsim\alpha_{\rm gr}$, where only GW radiation is present and $\mathcal{G}$ is given by Eq.~(\ref{eq:Ggr}), the constraint on the inclination angle is 
\begin{eqnarray}  \label{eq: iota_crit1}
    {\crd H(\iota)}&\lesssim& \left(\frac{5.5\times 10^{-6}}{c^2_{\Delta m}}\right)q^{1/2}\alpha^{7}\chi_s^{1/6}|\Delta m|^{1/6} f\left( e \right) ^{2/3} \left(1+\frac{\alpha^{15}}{\alpha^{15}_{\rm GW}}\right)\,,
\end{eqnarray}
where the cloud depletion due to GW emission of the cloud, as described in Eq.~(\ref{eq:Mc_GW_decay}), has been taken into account.
Given that $\alpha_{\rm gr}>\alpha_{\rm GW}$ as indicated in Tab.~\ref{tab:alpha}, the cloud mass $\bar{M}_c$ may be many orders of magnitude smaller than its saturated value, i.e. $\bar{M}_c\sim \alpha$. This can significantly reduce the backreaction and thus weaken the potential constraints on $\iota$. For $\alpha_{\rm gr}\gtrsim\alpha\gtrsim 
\alpha_{\rm NA}$, using the lower bound on $\mathcal{G}$ in Eq.~(\ref{eq:Gtpeak}) as a conservative estimate, the constraint becomes 
\begin{eqnarray}  \label{eq: iota_crit2}
   {\crd H(\iota)}&\lesssim& \left(\frac{0.04}{c^2_{\Delta m}}\right)q^{-1} \alpha^{-17}\chi_s^{-23/6}|\Delta m|^{-23/6}\,\bar{t}_{\rm peak}^{-3/2}\left(1+\frac{\alpha^{15}}{\alpha^{15}_{\rm GW}}\right)\,. 
\end{eqnarray}
Compared to Eq.~(\ref{eq: iota_crit1}), the constraint on $\iota$ shows a significantly stronger dependence on $\alpha$.

When the adiabatic conditions are met, it is important to examine the impact of resonance breaking effects. For the two decaying modes, the decay widths are approximately given by 
$\bar\Gamma_b\approx \frac{1}{24}\alpha^{10}$ for $\Delta m=-1$ and $\bar\Gamma_b\approx\frac{1}{24}\alpha^9(\alpha+\frac{1}{2}\chi_s)$ {\crd for} $\Delta m = -2$.
The critical value of {\crd initial state} population, $ \left| c_{a,1} \right|^2$, required for resonance breaking can be determined by substituting the expressions for $\mathcal{Z}$ and $\mathcal{B}$ into Eq.~(\ref{eq:Gammabreaking}). As before, we take different forms of $\mathcal{G}$ for $\alpha\gtrsim \alpha_{\rm gr}$ and $\alpha_{\rm gr}\gtrsim \alpha\gtrsim \alpha_{\rm NA}$, respectively, and the critical value is
\begin{eqnarray}\label{eq:Gammabreaking1}
    \begin{aligned}
        \left|c_{a,1} \right|^2&\approx \frac{\bar{\Gamma}_b}{2\mathcal{Z} \mathcal{B}}={\crd H(\iota)}^{-1}\left(1+\frac{\alpha^{15}}{\alpha_{\rm GW}^{15}}\right) \left(\alpha+\frac{(|\Delta m|-1)\chi_s}{2}\right)\\
        &\times\begin{cases}
	 \left(\frac{8\times 10^{-5}}{c^2_{\Delta m}}\right)\alpha ^5\chi_s ^{-5/3}|\Delta m|^{-5/3}f(e),\; \alpha \gtrsim \alpha _{\mathrm{gr}}\,,\\[1em]
	\left(\frac{0.03}{c^2_{\Delta m}}\right)q^{-1}\alpha ^{-11}\chi_s ^{-13/3}|\Delta m|^{-13/3}{\bar{t}_{\mathrm{peak}}}^{-1},\; \alpha _{\mathrm{NA}}\lesssim \alpha \lesssim \alpha _{\mathrm{gr}}\,.\\
\end{cases} 
    \end{aligned} 
\end{eqnarray} 
If the critical value $|c_{a,1}|^2$ could exceed one in some parameter space, it means that the transition has largely not occurred, and the cloud mass $\bar{M}_{c,1}$ remains at its saturated value.  

\begin{figure}[!ht]
\centering
\includegraphics[height=5.3cm]{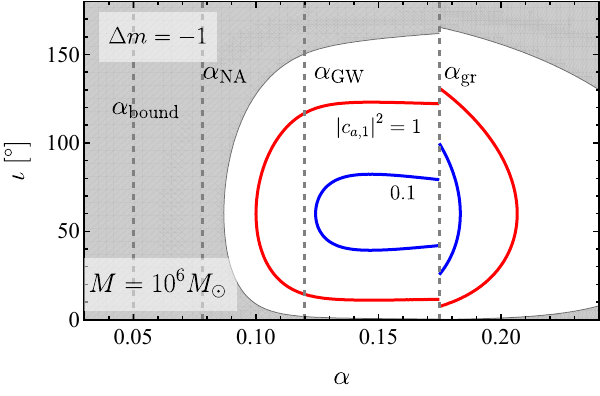}\quad
\includegraphics[height=5.3cm]{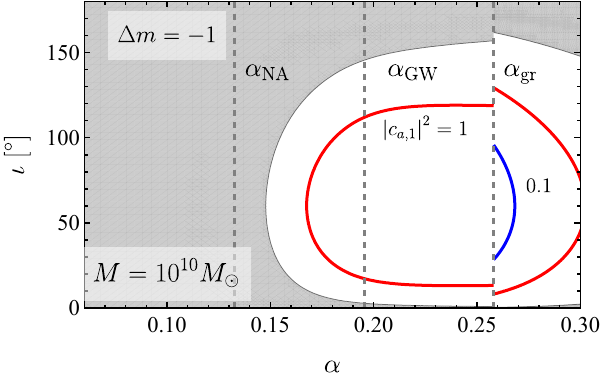}
\includegraphics[height=5.3cm]{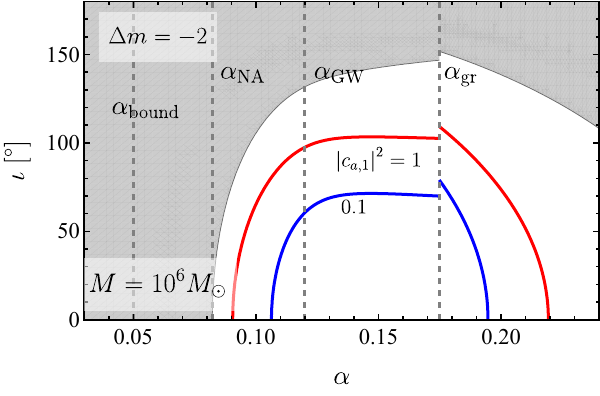}\quad
\includegraphics[height=5.3cm]{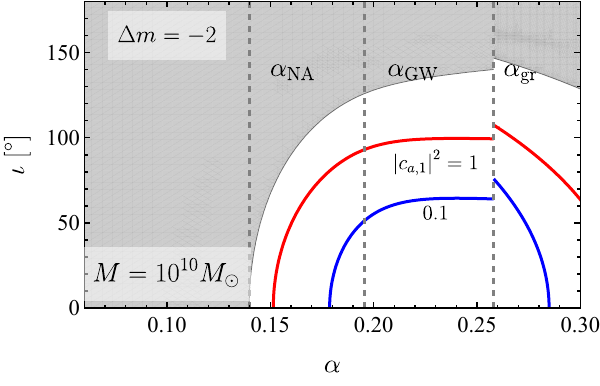}
\caption{Constraints for the {\crd initial state} $|a\rangle=\left|211\right>$ hyperfine transition on the plane of the gravitational fine structure constant $\alpha$ and the inclination angle $\iota$ for SMBHBs with a mass ratio $q\approx 0.1$, shown for $M=10^6M_{\odot}$ (left) and $M=10^{10} M_{\odot}$ (right), and $\Delta m=-1$ (upper) and $\Delta m=-2$ (lower). The shaded gray region represents the parameter space where the transition is non-adiabatic, with the boundary values determined by Eq.~(\ref{eq: iota_crit2}) for $\alpha_{\rm gr}\gtrsim\alpha\gtrsim 
\alpha_{\rm NA}$ and Eq.~(\ref{eq: iota_crit1}) for $\alpha\gtrsim \alpha_{\rm gr}$, respectively. In the adiabatic transition regime, the colored lines represent contours of the {\crd initial state} population at resonance breaking, $ \left| c_{a,1} \right|^2$ in Eq.~(\ref{eq:Gammabreaking1}). When $\alpha\gtrsim \alpha_{\rm GW}$, the cloud depletion due to its GW emission in Eq. \eqref{eq:Mc_GW_decay} becomes significant. The vertical gray lines mark various critical values of $\alpha$, as listed in Tab.~\ref{tab:alpha}.  
}
\label{orbit_evolution}
\end{figure}

Figure~\ref{orbit_evolution} summarizes the results and shows various constraints for the {\crd initial state} hyperfine transition as functions of $\alpha$ and  $\iota$. While there are some similarities to the findings of Ref.~\cite{Tomaselli:2024bdd} due to the same inclination angle  dependence through ${\crd H(\iota)}$, Fig.~\ref{orbit_evolution} reveals key differences in the hyperfine transition features for realistic SMBHBs. Firstly, the parameter space for non-adiabatic transitions, with boundaries defined by Eqs.~(\ref{eq: iota_crit1}) and (\ref{eq: iota_crit2}), has significantly expanded. For $\alpha\lesssim \alpha_{\rm gr}$, this expansion is due to the additional contributions to energy loss channels, particularly interactions with stars, before the binaries enter the GW radiation stage. For $\alpha\gtrsim \alpha_{\rm gr}$, it results from the inclusion of cloud depletion caused by GW emission of the cloud.
Secondly, in the adiabatic transition regime, the {\crd initial state} population at resonance breaking is greatly enhanced for the same reasons. Specifically, as shown, the cloud mass after the hyperfine transition $\bar{M}_{c,1}$ would not significantly fall below $0.1\alpha$. This greatly relaxes the constraint on $\iota$, previously identified in Ref.~\cite{Tomaselli:2024bdd}, regarding the necessity for nearly counter-rotating orbits to maintain sufficient cloud mass for future direct detection.
For all panels, the gradual change in curve slope around $\alpha_{\rm GW}$ is related to the considerable cloud depletion due to GW emission as described in Eq.(\ref{eq:Mc_GW_decay}). A discontinuity appears at $\alpha_{\rm gr}$ because the realistic orbital evolution time $t_{\rm evol}$ should be less than $t_{\rm peak}$, leading to a larger $\mathcal{G}$ in Eq.~\eqref{eq:Gtpeak} and thus a more stringent condition as one approaches $\alpha_{\rm gr}$ from the left. This discontinuity is significantly enhanced in cases where $\left|c_{a,1}\right|^2=0.1$, due to the high order of $\cos(\iota/2)$ in ${\crd H(\iota)}$.






As highlighted in Ref.~\cite{Tomaselli:2024bdd}, an efficient hyperfine transition can leave a distinct imprint on the orbital properties through the evolution of eccentricity $e$ and inclination $\iota$ during the floating orbit stage. 
For $\alpha \gtrsim \alpha_{\rm gr}$, where only GW radiation is relevant, 
the evolution tends to drive $\iota$ to zero and $e$ to a fixed point determined by the condition $2h(e)/f(e)=1$, i.e. $e\approx 0.46$. This fixed point serves as indirect evidence for the presence of a boson cloud in the binary’s history~\cite{Tomaselli:2024dbw}. 
However, as mentioned earlier, the adiabatic transition in this region might not be as efficient due to cloud depletion from GW emission.
As a result, the distinct imprint left on the orbital properties would likely be considerably weakened.
For $\alpha_{\rm gr} \gtrsim\alpha \gtrsim \alpha_{\rm NA}$,  the energy loss is dominated by other astrophysical process, where the angular momentum loss may exhibit behavior quite different from that induced by GW. Thus, the characteristic imprints on the orbital properties of binaries would also differ.

Finally, let us briefly comment on the fine transition (or Bohr transition). For the {\crd initial state}  $\left|a\right>=\left|211\right>$, the only fine transition is $\left|211\right>\to\left|200\right>$ with $\Delta\ell=-1$. According to the selection rule, the only nonzero contribution to the potential comes from the dipole mode with $\ell=1$, and thus the sum over $g$ is dominated by the $g=-1$ mode. Since the fine transition occurs at a smaller radius, with the semi-major axis  $\bar{a}_1\propto \alpha^{-10/3}$, additional environmental effects become significant only for smaller values of $\alpha$, i.e. $\bar{a}_1\gtrsim a_{\rm gr}$. In this regime, the transition may become inefficient, similar to the case of the hyperfine transition. Moreover, as explained in Ref.~\cite{Tomaselli:2024bdd}, even when the adiabatic condition is satisfied, the much larger decay width of the $|200\rangle$ state (i.e. $\bar{\Gamma}_b\propto \alpha^5$) makes the resonance-breaking condition easily achievable for fine transition. For Bohr transition, it has been argued that a floating orbit would not be possible due to the resonance breaking effects~\cite{Tomaselli:2024bdd}. Thus, we expect that both fine and Bohr transitions play a negligible role in the discussion of SMBHBs.

\subsection{Transition to unbound states}
\label{sec:ionization}

When the separation between the binary components is sufficiently small, the orbital frequency can become high enough to induce a transition from the {\crd initial state} to unbound continuum states by the potential in Eq.~(\ref{eq: tidal_V}). This process is referred to as the ionization of gravitational atoms~\cite{Baumann:2021fkf}. 
Given that GW radiation prior to the ionization significantly circularizes the orbit, we assume zero eccentricity for simplicity in the following discussion, where the semi-major axis $a$ reduces to $r$.  
This process is also characterized by the matrix element $\left< a \right|V(t) \left| K \right>$, with $|K\rangle$ a continuum state. For a circular inclined orbit, this matrix element can be decomposed into a sum over $g$ as shown in Eq.~(\ref{Fourier_coef}), with the expansion coefficients being of particular significance~\cite{Baumann:2021fkf,Tomaselli:2023ysb}:
\begin{equation} \label{eq: etaaK}
    \eta_{aK}^{(g)} (t) = -4\pi q\alpha \sum_{\ell ,m}\frac{d_{m,g}^{\ell}(\iota(t))}{2\ell +1}  Y_{\ell g}\left(\frac{\pi}{2},0\right)I'_{r}(t)I_{\Omega}\,.
\end{equation}
Here, the radial integration factor $I'_r(t)$ depends on the radial wave function of the continuum state and is given by:
\begin{eqnarray}\label{eq:Irp}
I'_{r}(t)=\int_0^{\infty}{\mu^{3/2} R_{n_a\ell _a}\left( r^{\prime} \right) R_{k;\ell _K}\left( r^{\prime} \right)F\left(r(t), r^{\prime} \right)  {r^{\prime}}^2\mathrm{d}r^{\prime}}\,.
\end{eqnarray}
For later discussion, it is useful to make explicit the scaling behavior of these quantities. By substituting $R_{n\ell}(r)$ and $R_{k;\ell}(r)$ from Eqs.~(\ref{eq: Rnl}) and (\ref{eq:Rkl}) into Eqs.~(\ref{eq:Irp}) and (\ref{eq: etaaK}), we find that
\begin{eqnarray}
   I'_{r}(t)=\frac{\alpha}{\sqrt{GM}} \bar{I}'_{r}(x,\bar{k}),\quad
   \eta_{aK}^{(g)} (t)&=& \frac{q\alpha^2}{\sqrt{GM}}\bar\eta_{aK}^{(g)} (x,\bar{k})\,,
\end{eqnarray}
where $\bar{I}'_{r}$ and $\bar\eta_{aK}^{(g)}$ are dimensionless functions of $x$ and $\bar{k}$ defined as
\begin{eqnarray}
    \bar{I}'_{r}(x,\bar{k})&=&\int_0^{\infty}\bar{R}_{n_a\ell_a}(x') \bar{R}_{\bar{k};\ell_K}(x'))\bar{F}(x,x')  x'^2 dx'\nonumber\\
    \bar\eta_{aK}^{(g)} (x,\bar{k})&=&-4\pi \sum_{\ell ,m}\frac{d_{m,g}^{\ell}(\iota(t))}{2\ell +1}  Y_{\ell g}\left(\frac{\pi}{2},0\right)\bar{I}'_{r}(x,\bar{k})I_{\Omega}\,.
\end{eqnarray}

The time evolution of a system with multiple bound states coupled to continuum states is inherently complex. However, the situation can be greatly simplified { in our parameter range of interest.}
Specifically, for $g\Omega$ sufficiently large to excite the unbound state and is far from the resonance condition for transitions between bound states, the coupling of the {\crd initial state} to other bound states can be neglected due to rapid oscillations { under the condition $q\alpha\ll1$ (see Appendix A.3 in Ref.~\cite{Baumann:2021fkf} for details). This condition is readily satisfied for $\alpha\lesssim 0.3$ and $q\sim 0.01-0.1$.}
By ignoring other bound states, the evolution simplifies to a generalized form of the two-state Landau-Zener transition, which can be effectively described by~\cite{Baumann:2021fkf}  
\begin{subequations}
\begin{align}
    \mathrm{i}\frac{\mathrm{d}c_a}{\mathrm{d}\bar{t}}&=q\alpha^4\int {\bar{\eta}_{Ka}}^{(g)*}\left( \bar{t} \right) \mathrm{e}^{\mathrm{i}\left( g\bar{\Omega}-\Delta \bar{E} \right) \bar{t}}\bar{c}_k\mathrm{d}\bar{k}\,,\label{eq: ca_Sch_2}
\\
\mathrm{i}\frac{\mathrm{d}\bar{c}_k}{\mathrm{d}\bar{t}}&=q\alpha^2 \bar{\eta}^{(g)}_{Ka}\left( \bar{t} \right) \mathrm{e}^{-\mathrm{i}\left( g\bar{\Omega}-\Delta \bar{E} \right) \bar{t}}c_a\,,\label{eq: ck_Sch_2}
\end{align}
\end{subequations}
where 
$\Delta \bar{E}=GM(E_K-E_a)$ and $\bar{c}_k=c_k/\sqrt{GM}$.
By integrating out the continuum states, the Schrödinger equation for the bound state simplifies to $\mathrm{i}\,\mathrm{d}c_a/\mathrm{d}\bar {t}=\bar{\mathcal{E}}_a c_a$, where $\mathcal{E}_a$ is the so-called induced energy, which accounts for the integration over the continuum states. The evolution of the occupation probability of the bound state is then determined by $\text{Im} \mathcal{E}_a$ and is dominated by the states resonating with the system, i.e.
\begin{align} \label{eq: dca2_dt}
    \frac{\text{d}\ln|c_a(\bar{t})|^2}{\text{d}\bar{t}} 
    \approx 
    - q^2\alpha^3\sum_{\ell_K,g}\frac{|\bar{\eta}^{(g)}_{Ka}(x)|^2}{\bar{k}^{(g)}(x)}\Theta(\bar{k}^{(g)}(x)^2)
    \equiv -q^2\alpha^3 \mathcal{D}(x)\,.
\end{align}
Here, $k^{(g)}(t)=\sqrt{ (g\Omega(t)+\epsilon_a+\mu)^2-\mu^2}$
denotes the wavenumber at the resonant frequency. The $\bar{t}$ dependence is encoded by the explicit $x(\bar{t})$ dependence of the dimensionless wavenumber $\bar{k}^{(g)}$ and $\bar\eta_{aK}^{(g)}$, and their explicit expressions are given as\footnote{Note that, in principle, an additional term $\alpha^2(gx^{-3/2}-(2n_a^2)^{-1})^2$ should be present in the square root for $\bar{k}^{(g)}(x)$. However, this term is highly suppressed by $\alpha^2$ for small $\alpha$, making it important only at small values of $x$, which will not be relevant for our later discussion.} 
\begin{eqnarray}
\bar{k}^{(g)}(x)=\frac{GM}{\alpha^{2}} k^{(g)}(x)\approx\sqrt{2\left(g x^{-3/2}-(2n_a^2)^{-1}\right)}\,,\quad
\bar\eta_{aK}^{(g)}(x)=\bar\eta_{aK}^{(g)}(x,\bar{k}^{(g)}(x))\,,
\end{eqnarray}
where $GM \epsilon_a=-\alpha^3/(2n_a^2)$ is the binding energy of the $|a\rangle$ state from Eq.~(\ref{eq: energy_level}). 
The Heaviside Theta function implies that when the distance $x$ is large, the {\crd initial state} can only transition to the continuum states with a high overtone, and a $g$-th mode starts to contribute as the binary system crosses the critical distance $x_g=(2n_a^2g)^{2/3}$. In the final expression of Eq.~(\ref{eq: dca2_dt}), all $x$-dependent functions are incorporated into the form factor $\mathcal{D}(x)$ for the convenience of later discussion.  
This process follows the same selection rules as transitions to bound states.

During ionization, the energy of the boson cloud is transferred to infinity at a rate referred to as the ionization power, i.e. $ P_{\mathrm{ion}}\left( t \right) \equiv \text{d}E_{\rm ion}/\text{d}t$. Here, $E_{\rm ion}$ denotes the total energy contained within the continuum relative to the initial bound state, and the power is proportional to the temporal evolution of the occupation probability of the continuous states. As for Eq.~(\ref{eq: dca2_dt}), the resonant modes dominate the contribution, and the dimensionless ionization power $\bar{P}_{\mathrm{ion}}(t) \equiv d (E_{\rm ion}/M)/d\bar{t}$ can be expressed as~\cite{Baumann:2021fkf} 
\begin{eqnarray}  \label{eq: P_ion}
  \bar{P}_{\mathrm{ion}}(t) 
  &\approx& G\, \frac{M_c}{\mu}|c_a(t)|^2\sum_{\ell_K, g} g\Omega(t) \frac{\mu |\eta^{(g)}_{Ka}|^2}{k^{(g)}(t)}\Theta(k^{(g)}(t)^2)\nonumber\\
  &\approx&  q^2\alpha^5\bar{M}_c(t)\mathcal{F}_{\rm ion}(x(\bar{t}))\,,
\end{eqnarray}
where $\bar{M}_c(\bar{t})=M_c|c_a(\bar{t})|^2/M$ is the time-dependent cloud mass normalized by the BH mass. In the second line, we make explicit the scaling behavior of the power, where the dimensionless form factor $\mathcal{F}_{\rm ion}$ encodes the non-trivial $x$-dependence of the ionization power, as defined by
\begin{eqnarray}\label{eq:Fion}
    \mathcal{F}_{\rm ion}(x)\equiv \sum_{\ell_K,g}g x^{-3/2}\frac{ |\bar{\eta}^{(g)}_{Ka}(x)|^2}{\bar{k}^{(g)}(x)}\Theta(x_g-x)\,.
\end{eqnarray}
where $\Theta(\bar{k}^{(g)}(x)^2)$ is replaced by $\Theta(x_g-x)$. 
Note that with $\eta_{Ka}^{(g)}$ defined in Eq.~(\ref{eq: etaaK}), the ionization form factor still depends on the inclination angle $\iota$ through the Wigner $d$-matrix $d_{m,g}^{\ell}$. However, unlike in the cases of hyperfine transitions, $\mathcal{F}_{\rm ion}$ receives contributions from a larger number of continuous states. Consequently, the summation over $m$ reduces the influence of the Wigner $d$-matrix to approximately $0.5\sim 1$, making the $\iota$-dependence a minor effect~\cite{Tomaselli:2023ysb}. 

\begin{figure}[h]
    \centering
    \includegraphics[height=6cm]{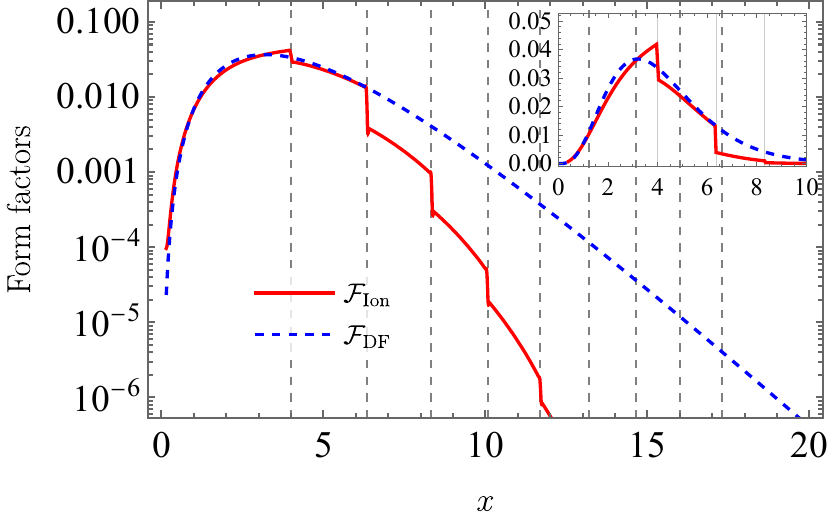}
    \caption{The form factors $\mathcal{F}_{\rm ion}$ in Eq.~(\ref{eq:Fion}) and $\mathcal{F}_{\rm DF}$ in Eq.~(\ref{eq:FormDF}) as functions of $x$ for a counter-rotating orbit with $\iota=\pi$. The inset provides a comparison on a linear scale.}
    \label{fig: FF1}
\end{figure}

Figure~\ref{fig: FF1} displays the form factor of the ionization power for the $|211\rangle$ state. Here, $\eta_{Ka}^{(g)}$ is evaluated by using Ref.~\cite{Baumann:2021fkf}.  A key feature is the discontinuity of the power at the critical radius $x_g$, where the resonance condition is barely satisfied. This comes from the finite contribution of the continuous states to the power in small wavenumber limit. Specifically, as $\bar{k}\to 0$, $|\bar\eta^{(g)}_{Ka}|^2\propto \bar{k}^{(g)}$ as shown in Eq.~(\ref{eq:Rklsmallk}), which precisely cancels out with $\bar{k}^{(g)}$ in the denominator of Eq.~(\ref{eq:Fion}).    


The ionization power turns out to show some similarity to the dynamical friction power $P_{\rm DF}$ as noted in Ref.~\cite{Tomaselli:2023ysb}.  When considering the boson cloud as a fluid, the BH companion induces density perturbations, or ``wakes'', as it travels through the cloud.  These wakes create a gravitational pull on the companion, expressed as~\cite{Chandrasekhar:1943ys, Ostriker:1998fa, Hui:2016ltb, Zhang:2019eid} 
\begin{equation}\label{eq:FDF}
    F_{\rm DF}=-\frac{4\pi(G M q)^2}{v_{\rm rel}^2}\rho_a(\mathbf{r}) \mathcal{C}(\beta, \mu v_{\rm rel} r)\,, 
\end{equation}
where the orbital radius $r$ serves as a cut-off for the computation. Here, $v_{\rm rel}\approx 1/\sqrt{\bar{r}}$ is the relative velocity between the cloud and the companion,\footnote{Given that the cloud's rotational speed is much slower than the companion's orbital speed, we have $v_{\rm rel}\approx 1/\sqrt{\bar{r}}$ from the Keplerian relation.} {\crd $\bar{r}\equiv r/GM$ is the dimensionless radius,} and $\beta\equiv q\alpha/v_{\rm rel}\approx q\sqrt{x}$. In the particle limit, i.e. $\beta\gg 1$, the form factor $\mathcal{C}$ depends only on the ratio $\Lambda=\mu v_{\rm rel} r/\beta\approx 1/q\gg1$, with $\mathcal{C}(\Lambda)\approx \log \Lambda$, reproducing the classical Chandrasekhar result~\cite{Chandrasekhar:1943ys}. In the wave limit, $\beta\ll 1$, $\mathcal{C}$ is only a function of the product $\mu v_{\rm rel} r\approx \sqrt{x}$, with $\mathcal{C}(x)\approx \mathrm{ci}\left( 2\sqrt{x} \right) +\mathrm{sinc}\left( 2\sqrt{x} \right) -1$, where the cosine integral function $\mathrm{ci} (z)\equiv \int_0^z(1-\cos t) \text{d}t/t$ and $ \mathrm{sinc}(z)\equiv \sin z/z$~\cite{Hui:2016ltb}. 
For SMBHBs with $q\sim 0.01-0.1$ and $x\sim\mathcal{O}(10)$, we have $\beta\lesssim 1$, making the wave limit a suitable approximation. 
The dynamical friction force, $F_{\rm DF}$, results in energy emission power given by $P_{\rm DF}\equiv F_{\rm DF}v_{\rm rel}$.
 By substituting $\rho_a(\mathbf{r})$ from Eq.~(\ref{eq:rhoa}) into Eq.~(\ref{eq:FDF}), we then obtain 
\begin{equation}  \label{eq: P_DF}
  \bar{P}_{\mathrm{DF}}(t)
  =G P_{\rm DF}(\bar{t})
  \approx q^2\alpha ^{5}\bar{M_c}(\bar{t})\,\mathcal{F}_{\rm DF} \left( x(\bar{t}) \right)\,.
\end{equation}
It shows the same scaling as $\bar{P}_{\rm ion}$ in Eq.~(\ref{eq: P_ion}), but with a different form factor
\begin{equation}\label{eq:FormDF}
   \mathcal{F}_{\rm DF}(x)\equiv 4\pi x^{1/2}\left|\bar{R}_{n_a\ell_a}(x)\right|^2\mathcal{C}(x) \left|\sum_{m'}d^{(\ell_a)}_{m_a m'}(\iota)Y_{\ell_a m'}(\pi/2,0)\right|^2\,.
\end{equation}
The final term captures the dependence on the inclination angle. In the case of the  $\left|211\right>$ state, this factor simplifies to $1-\sin(\iota)/\sqrt{2}$, resulting in only about a 10\% variation with respect to $\iota$. As shown in Fig.~\ref{fig: FF1}, $\mathcal{F}_{\rm DF}(x)$ behaves similarly to $\mathcal{F}_{\rm ion}(x)$ at relatively small values of $x$. However, for large $x$, we find that $\mathcal{F}_{\rm ion}(x)\propto \exp(-x^{1.3})$, which decays more rapidly than  $\mathcal{F}_{\rm DF}(x)\propto \exp(-x)$. In fact, at even larger values of $x$, where the particle limit applies with $\beta\gg1$, the form factor introduces additional $q$ dependence through $\mathcal{C}\approx \log(1/q)$. This reveals the intrinsic difference between ionization and dynamical friction.   

As a side note, the boson cloud could accrete onto the companion BH, providing an additional channel for energy loss. The accretion power is detailed in Ref.~\cite{Baumann:2021fkf}, and its dimensionless counterpart can be expressed as
\begin{eqnarray}
    \bar{P}_{\text{acc}}(t) = q^2\alpha^8 \bar{M}_c(t)\mathcal{F}_{\text{acc}}(x(\bar{t}))\,,
\end{eqnarray}
where $\mathcal{F}_{\text{acc}}(x)$ denotes the corresponding form factor. 
Compared to Eq.~(\ref{eq: P_ion}), $\bar{P}_{\text{acc}}$ has a different $\alpha$ scaling, being suppressed by an additional factor of $\alpha^3$ relative to $\bar{P}_{\rm ion}$.
Therefore, even though $\mathcal{F}_{\rm acc}(x)$ may be larger than $\mathcal{F}_{\rm ion}(x)$ at both small and large $x$, for $\alpha\lesssim 0.2$ and $3\lesssim x\lesssim 10$, the accretion power remains subdominant and is thus ignored in subsequent discussions.

To analyze the backreaction on the orbit, we need to consider the simultaneous evolution of the binary orbit and boson cloud. By applying energy conservation, i.e. $\text{d} E_{\rm orb}/\text{d}t=-P_{\rm GW}-P_{\rm ion}$ {\crd (with the orbital energy of the companion SMBH: $E_{\rm orb}\equiv -qGM^2/(2r)$)}, and substituting into Eqs.\eqref{eq: dca2_dt} and (\ref{eq: P_ion}), the time evolution of the orbit and the cloud mass are given by~\cite{Baumann:2021fkf}
\begin{subequations}\label{eq:xtMct}
\begin{align}
 \frac{\mathrm{d} x} {\mathrm{d}\bar{t} }&=-2q^{-1}x^2\alpha^{-2}\bar{P}_{\mathrm{GW}}[1+\mathcal{R}(x)],  
\\
\frac{\mathrm{d} \bar{M}_c}{\mathrm{d} \bar{t}}&=- q^2\alpha^3 \bar{M}_{c} \mathcal{D}(x)\,,
\end{align}
\end{subequations}
where $\bar{P}_{\rm GW}(x)=\frac{32}{5}q^2 x^{-5} \alpha^{10}$ denotes the power of GW radiation and $\mathcal{R}(x)\equiv \bar{P}_{\rm ion}(x)/\bar{P}_{\rm GW}(x)$ denotes the ratio of the two power:
\begin{equation}\label{eq:Rx}
    \mathcal{R}(x)\equiv \frac{5}{32}\alpha^{-5} \bar{M}_c(x)x^5\mathcal{F} _{\mathrm{ion}}(x)\,.
\end{equation}
If cloud depletion is not considered, the ratio $\mathcal{R}$ is significantly enhanced for small values of $\alpha$, as highlighted in Ref.~\cite{Baumann:2021fkf, Zhang:2019eid}. However, as we will demonstrate below, cloud depletion can have a strong impact on the maximum values of $\mathcal{R}$. To solve the cloud mass evolution, note that the right-hand sides of Eq.~(\ref{eq:xtMct}) consist entirely of functions of $x$. Thus, it is convenient to solve for $\bar{M}_c$ as a function of $x$ by dividing the two equations: 
\begin{eqnarray}  \label{eq: Mc}
\begin{aligned}
    \frac{\mathrm{d}\bar{M}_c}{\mathrm{d}x}=\dfrac{5q\bar{M}_{c}x^{3}\mathcal{D}(x)}{64\alpha ^5\left[ 1+\mathcal{R}(x)\right]}\,.
\end{aligned}
\end{eqnarray} 
This equation can only be solved numerically given the complex $x$-dependence in the form factors. However, to understand its dependence on $q$, $\alpha$ and $\bar{M}_{c,1}$, it is useful to examine the semi-analytical results under certain conditions. Specifically, when the ionization power dominates over the GW power, i.e. in the $\mathcal{R}\gg 1$ limit, Eq.~(\ref{eq: Mc}) can be approximated as $\text{d}\bar{M}_c/\text{d}x \approx q\mathcal{D}(x)/(2x^2\mathcal{F}_{\rm ion}(x))$, where the explicit $\alpha$-dependence cancels out. 
The change in cloud mass from some initial value, $\Delta \bar{M}_c(x)\equiv \bar{M}_{c}(x)-\bar{M}_{c}(x_2)$, is then given by
\begin{align}  \label{eq: dxMc_largeR}
    \frac{\Delta \bar{M}_c(x)}{\bar{M}_c(x_2)}\approx 1- \frac{q} {\bar{M}_c(x_2)}I(x)\,,
\end{align}
where $I(x) = \int^{x_2}_x\text{d}x'  \frac{1}{2x'^2}\frac{\mathcal{D}(x')}{\mathcal{F}_{\rm ion}(x')}$. With all $x$-dependence encoded in $I(x)$, this indicates that the relative cloud mass change in this regime is only proportional to $q/\bar{M}_c(x_2)$. 
When GW radiation power dominates, i.e. $\mathcal{R}\ll1$, the cloud mass undergoes exponential decay, as 
\begin{equation}  \label{eq: small_R}
    \bar{M}_c(x)=\bar{M}_{c}(x_2) \exp\left(-\frac{5}{64}q\alpha ^{-5}\int_x^{x_2}  \text{d}x'{x'}^3 \mathcal{D}(x')\right)\,.
\end{equation}
Although this decay shows a strong dependence on $\alpha$ (i.e. with the exponential term being enhanced by $\alpha^{-5}$ for small $\alpha$), the potential cloud depletion due to GW emission remains negligible up to $x \sim 15$ because of the rapid decay of $\mathcal{D}(x) \propto \exp(-x^{1.3})$ at large $x$.\footnote{For $x\gtrsim 15$, the relative change of cloud mass is proportional to $q\alpha^{-5}\int_x^{x_2}dx'x'^3\mathcal{D}(x')\approx q\alpha^{-5} 10^{-16}\ll 1$ for $q<0.1$ and $\alpha >0.01$, which makes it negligible within the parameter space of interest.} Thus, the initial value of the cloud mass in Eq.~(\ref{eq: dxMc_largeR}) is approximately given by $\bar{M}_{c,2}\approx \bar{M}_{c,1}$.    

\begin{figure}[ht]
    \centering
    \includegraphics[height=5.3cm]{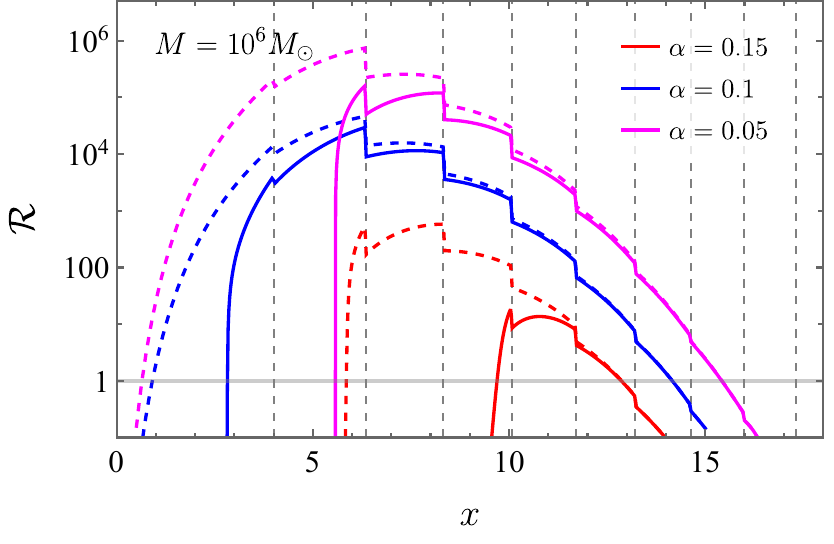}
    \includegraphics[height=5.3cm]{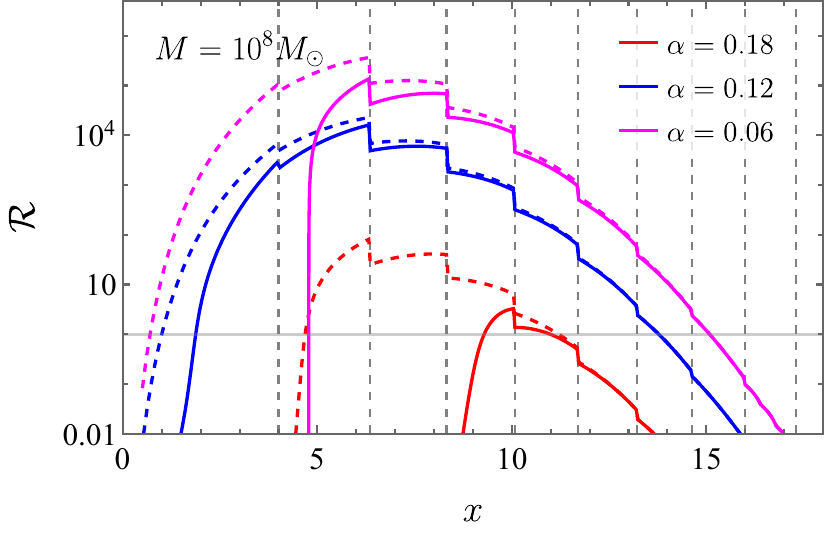}
    \includegraphics[height=5.3cm]{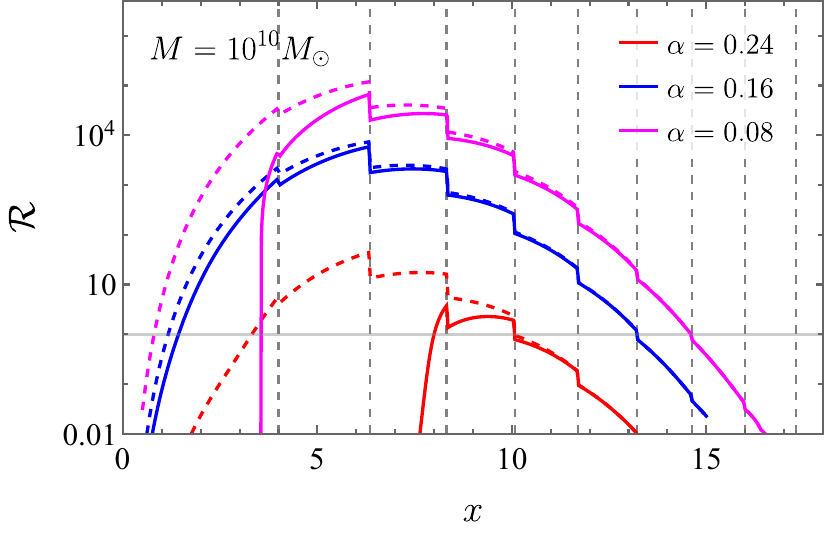}
    \caption{Numerical results of ionization-to-GW power ratio $\mathcal{R}$ in Eq.~(\ref{eq:Rx}) as a function of $x$, taking into account cloud depletion, for the BH mass $M=10^6M_{\odot}$ (upper left),  $M=10^{8}M_{\odot}$ (upper right), and $M=10^{10}M_{\odot}$ (lower), respectively.  Different colored lines represent various benchmark values of $\alpha$, chosen according to Tab.~\ref{tab:alpha}. Solid lines correspond to the mass ratio $q=0.1$, while dashed lines correspond to $q=0.01$. 
    }
    \label{fig: ratio_left}
\end{figure}

Figure~\ref{fig: ratio_left} displays the numerical results of the ionization-to-GW power ratio, taking into account the numerical evolution of cloud mass described in Eq.~(\ref{eq: Mc}), for selected benchmark values of the SMBHB's total mass $M$, mass ratio $q$ and $\alpha$, with consideration of the critical $\alpha$ values listed in Tab.~\ref{tab:alpha}. For each BH mass, we examine two typical values of $\alpha<\alpha_{\rm GW}$ (represented by the magenta and blue curves), where the inefficiency of the hyperfine transition and GW emission allows the cloud mass at the ionization stage to remain at its saturated value, i.e. $\bar{M}_{c,1}\approx \alpha$. In these cases, cloud depletion becomes significant only at relatively small $x$. For a given $\alpha$, a significant cloud depletion occurs earlier (i.e. a larger $x$) for the $q=0.1$ case compared to the $q=0.01$ case, consistent with the $q$-dependence of the relative change in cloud mass described in Eq.~\eqref{eq: dxMc_largeR}. Similarly, for a given $q$, cloud depletion happens sooner for smaller $\alpha$, since the relative change in Eq.~\eqref{eq: dxMc_largeR} is proportional to $\bar{M}_c(x_2)^{-1}\approx \alpha^{-1}$. We also display the results for a scenario with larger $\alpha$ values, where $\alpha>\alpha_{\rm GW}$, and the GW emission as described in Eq.~\eqref{eq:Mc_GW_decay} is not negligible. This can cause cloud depletion to occur even earlier, significantly suppressing the peak value of $\mathcal{R}$ compared to the one without depletion. Comparing the three panels, cloud depletion is less efficient in the more massive SMBHB case. This is mainly because the relevant parameter range of $\alpha$ shifts to higher values, resulting in weaker suppression related to $\bar{M}_{c,2}$. Overall, considering the complex dependence of $\bar{M}_{c,1}$ on $\alpha$, we can conclude that ionization effects are more detectable for SMBHBs with relatively small mass ratios $q$ and intermediate $\alpha$ values.

\section{Potential constraints from observations of SMBHBs}
\label{sec:observation}

\subsection{{Electromagnetic observations}}

Identifying { gravitationally bound SMBHBs at parsec to sub-parsec separations} via electromagnetic observations poses significant challenges, due to their small angular separations on the sky and the uncertainties concerning the uniqueness of their observational signatures. 
Several techniques are currently used to search for such systems (see Ref.~\cite{DeRosa:2019myq} for a review): (a) photometric variability in quasar light curves, which may indicate the orbital motion of a binary; (b) spectroscopic methods, such as broad emission-line velocity shift or peculiar broad emission-line ratios;
and (c) radio imaging searches with the very long baseline interferometry, which allows for resolving close binary structures in radio-loud galaxies. Over the past few decades, these methods have led to tens of SMBHB candidates~\cite{2015Natur.518...74G, Valtonen:2016awd, 2010ApJ...724L.166I, 2016ApJ...822....4L, Yan:2015mya, 2012ApJ...759..118B, 2019ApJS..241...33L, Rodriguez:2006th, 2017ApJ...843...14B, 2014ApJ...786..103L}, though their exact nature remains a topic of debate within the astronomical community.
Generally, photometric variability methods typically target SMBHBs {\crd with orbital periods of} $T\sim \mathcal{O}(1)-\mathcal{O}(10)\,$yr. Spectroscopic searches focus on binaries with dimensionless separations in a narrow range around $\bar{r}\sim 10^{4}$ {(i.e. $r\sim 10^4 GM$)}~\cite{2010ApJ...725..249S}. Radio imaging can resolve binaries with separations $r$ down to the pc scale, but likely not much smaller, thereby targeting binaries at a relatively early inspiral stage. These methods collectively provide a potential range of SMBHB properties that are suitable for electromagnetic observations.



Before discussing the direct detection of ionization effects through electromagnetic observations, let's briefly comment on the role of gas accretion on SMBH, as such observations often depend on highly luminous active galactic nuclei (AGN). While the majority of SMBHs exist in a quiescent state with extremely low accretion rates, a small fraction operate as AGN, characterized by high accretion rates that release intense energy. 
SMBHs undergo significant gas accretion only during a small portion of their total growth time—a phase known as the duty cycle~\cite{Kormendy:2013dxa}. The mass increase from accretion can be roughly estimated as: $\Delta M_{\rm acc}/M\sim f_{\rm edd}\Delta t_{\rm acc}/\tau_s$, where $f_{\rm edd}$ is the Eddington ratio
during significant accretion,  $\Delta t_{\rm acc}$ is the duration of high accretion, and $\tau_s\sim 4.5\times 10^7\,$yr is the {\crd Salpeter} time scale. 
{ The accretion rate of AGNs has been observed to differ between those located in elliptical and spiral galaxies. For AGNs in elliptical galaxies, which typically have limited gas supplies due to their late evolutionary stage,} simulations suggest that SMBHs with $ M\sim 10^6$-$10^9 M_\odot$ can periodically experience high accretion rates of $f_{\rm edd}\sim 0.1$ with a { low} duty cycle of $~10^{-4}$~\cite{Yoon:2018jya,Yuan:2017ylc}. Taking $t_{\rm peak}\sim 5\,$Gyr as an upper limit for the relevant time scale for the binary, the accretion duration 
$\Delta t_{\rm acc}\sim 10^{6}\,$yr, resulting in a relative mass change $\Delta M_{\rm acc}/M$ of less than $1\%$. 
{ In contrast,} AGNs in active spiral galaxies { can maintain accretion rates of $f_{\text{edd}} \sim 0.01-0.1$ with a higher duty cycle (up to $10^{-2}$)}~\cite{1981Natur.294..427S},
due to enhanced gas transport from various processes~\cite{Abdulrahman_2022}.
{As a result, even under extreme accretion cases, $\Delta M_{\rm acc}/M$ is only about 10\%, which remains a relatively modest change.}
Therefore, although SMBHB candidates identified through electromagnetic observations may currently exhibit high accretion rates, significant mass { growth during the binary's evolution is not expected.} 
This ensures that the theoretical framework { discussed in Sec.~\ref{sec:theory}} remains applicable. 
It is noteworthy that if a boson cloud forms around the central SMBH in an AGN today, the high accretion rate would result in significant time-evolution of $\alpha$ and associated new phenomena~\cite{Sarmah:2024nst}. Here, we focus on the cloud generated in earlier times and its observational consequences.

For detection methods sensitive to orbital periods, a key observable for ionization effects is the orbital period decay rate, $\dot{T}$, {\crd with $T\equiv 2\pi/\Omega$}. This rate is proportional to the orbital energy loss rate with
\begin{equation} \label{eq:T_dot}
\dot{T}=\frac{\mathrm{d}T}{\mathrm{d}E_{\rm orb}}\frac{\mathrm{d}E_{\rm orb}}{\mathrm{d}t}
=-\frac{192\pi}{5}q\bar{r}^{-5/2} (1+\mathcal{R}),
\end{equation}
where we have already applied the explicit forms of $P_{\rm GW}$ and $P_{\rm ion}$ in Eq.~(\ref{eq:xtMct}). In the regime that the ionization power dominates over the GW one, i.e. $\mathcal{R}\gg 1$, the orbital period decay rate can be expressed conveniently as a function of $x$, 
\begin{eqnarray}\label{eq:Tdot}
    \dot{T}\approx -6\pi q \bar{M}_c(x) x^{5/2}  \mathcal{F}_{\rm ion}(x)\,.
\end{eqnarray}
Notably, the $\alpha^{-5}$ enhancement seen in the ratio $\mathcal{R}$, as shown in Eq.~(\ref{eq:Rx}), is exactly canceled out by the $\alpha$ dependence in $\bar{r}$ when expressed as a function of $x$. As a result, $\dot{T}$ shows no explicit dependence on $\alpha$. 
In the optimal case that the {\crd cloud depletion is negligible}, the $x$ dependence of $\dot{T}$ is determined solely by $x^{5/2}\mathcal{F}_{\rm ion}(x)$, which can reaches the maximum around 1.3 when $4\lesssim x\lesssim 6$ from Fig.~\ref{fig: FF1}. This provides an upper bound for $|\dot{T}|$, i.e. $|\dot{T}|\lesssim 50\%$ for $q\lesssim 0.1$ and $\bar{M}_c(x)\lesssim \alpha\approx 0.2$ from Eq.~(\ref{eq:Tdot}), showing that it can reach up to an order of magnitude of $\mathcal{O}(10\%)$.  

\begin{figure}
    \centering
    \includegraphics[height=6.2cm]{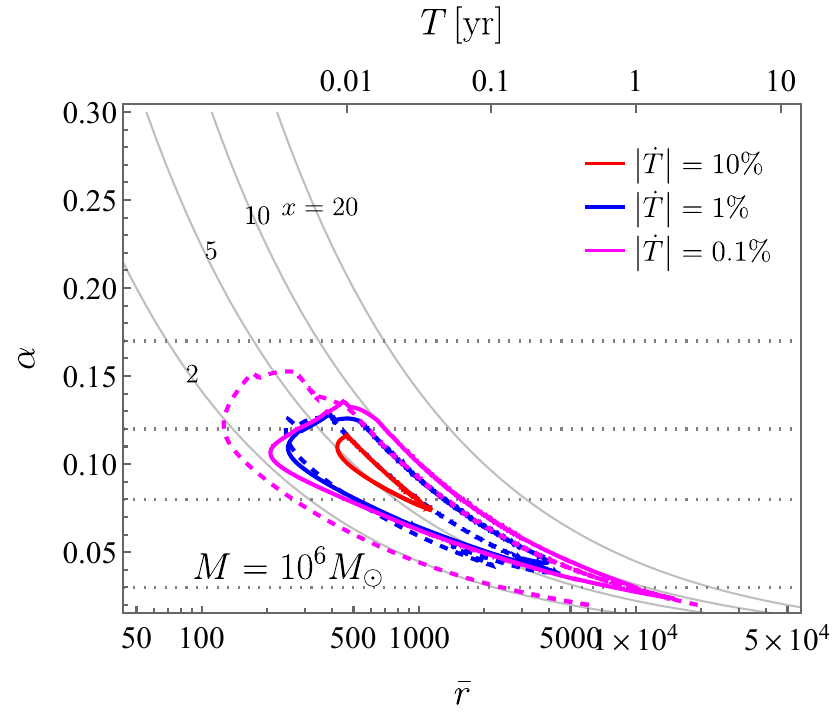}
    \includegraphics[height=6.2cm]{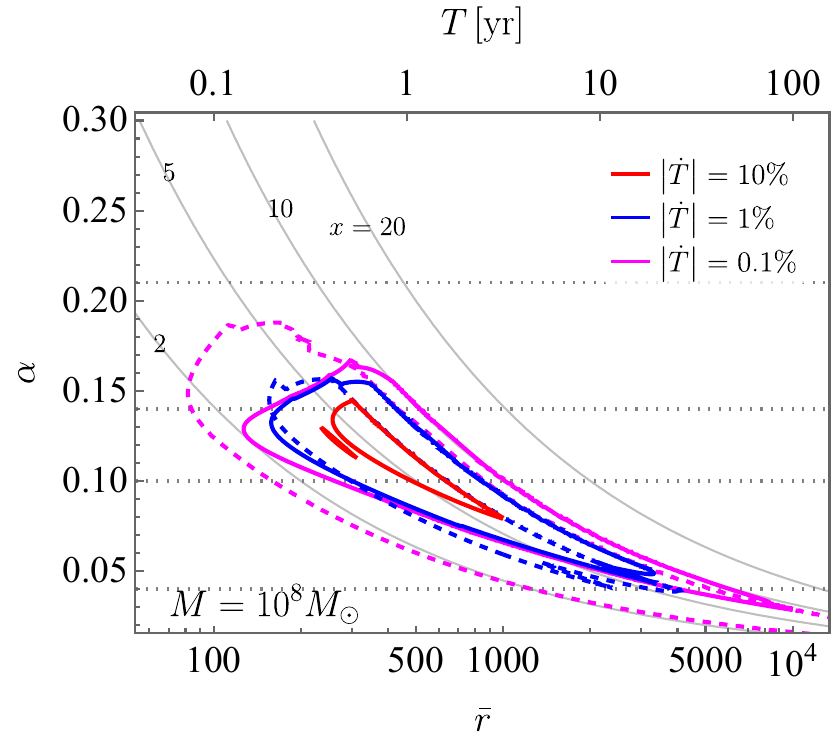}\quad
    \includegraphics[height=6.2cm]{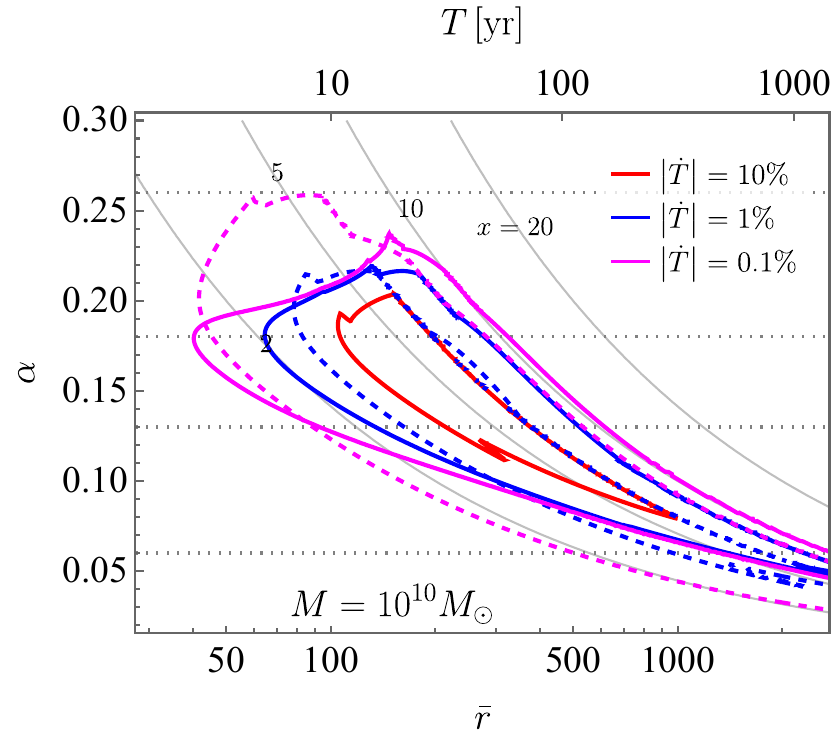}
    \caption{Orbital period decay rate $|\dot{T}|$ as a function of $\bar{r}$ (or $T$) and $\alpha$, accounting for cloud depletion in Eq.~(\ref{eq:Tdot}) whenever $\mathcal{R}\gg1$, for the BH mass {\crd $M=10^6M_\odot$ (upper left), $M=10^8 M_{\odot}$ (upper right) and $M=10^{10} M_{\odot}$ (below). } In {\crd all} panels, different colored lines represent { contours of different $|\dot{T}|$ values.} 
    Solid lines correspond to the mass ratio $q=0.1$ and dashed lines to $q=0.01$.  
    The gray curves indicate values of $x= \alpha^2\bar{r}$. The horizontal gray dashed lines denote $\alpha_{\rm inst}, \alpha_{\rm NA}, \alpha_{\rm GW}$ and $\alpha_{\rm gr}$, from bottom to top for different masses, as shown in Tab.~\ref{tab:alpha}. 
    }
    \label{fig:T_dot}
\end{figure}

To establish a more direct connection to the observations, Fig.~\ref{fig:T_dot} presents the contours of { different $|\dot{T}|$ values}
on the $\bar{r}(T)-\alpha$ plane, taking into account the cloud  depletion in $\bar{M}_c(x)$ discussed earlier. 
Here, we set the minimal { contour level} of $|\dot{T}|$ to be 0.1\%. Achieving { this level of precision in measuring orbital decay}
is quite challenging for electromagnetic observations of SMBHBs. 
However, for the well-known SMBHB candidate OJ 287, the original model suggests that the orbital period decay rate could indeed be measured with an accuracy as high as 0.1\%~\cite{Dey:2018mjg}. 
Fig.~\ref{fig:T_dot} indicates that the maximum allowed $|\dot{T}|$ occurs around $x\sim 5$ within the relevant range of $\bar{r}$. It can potentially exceed $1\%$ and $10\%$ for the $q=0.01$ and $q=0.1$ cases, respectively. 
For a given BH mass $M$, the values of $|\dot{T}|/q$ for $q=0.1$ and 0.01 overlap at upper boundary due to negligible cloud depletion effects. However, at the lower boundaries with smaller values of $\alpha$, cloud depletion becomes quite significant for the $q=0.1$ case. Consequently, ionization power dominates over a considerably smaller parameter space, resulting in reduced contours for the $q=0.1$ case.  
Comparing { different} BH mass cases, the parameter space expands for more massive BH cases in larger $\alpha$ and smaller $\bar{r}$ regime. This occurs because, as the BH mass increases, the relevant range for $\alpha$ shifts to higher values, resulting in inefficient cloud depletion in the expanded parameter region. 
{ Observationally, the predicted theoretical regime aligns well with the parameter ranges preferred for the photometric variability method.}
Specifically, for the $M=10^{10}M_{\odot}$ case,  the region with the maximal decay rate coincides with the area where $T\sim \mathcal{O}(10)$yr,  making it ideally suited for 
{ detection via photometric variability.}
For the $M=10^{8}M_{\odot}$ case, the maximal decay rate regime corresponds to $\bar{r}\sim \mathcal{O}(100)$ and $T\sim 1$yr, which can also be probed by photometric variability.
{ In the $M=10^{6}M_{\odot}$ case, the orbital decay rate peaks at a relatively short period of $T\sim \mathcal{O}(1)$day, requiring high-cadence sampling to be detected through photometric monitoring.
For the other two methods, the regime around $\bar{r}\sim 10^{4}$ can in principle be explored spectroscopically, but}
is less compelling as $\alpha$ mostly falls below $\alpha_{\rm inst}$ in this area. {\crd Radio imaging appears less relevant here, because the theoretical prediction for $|\dot{T}|$ at $r\sim \mathcal{O}(1)\,$pc, corresponding to $\bar{r}\approx 2\times 10^3 (10^{10}M_\odot/M)$, remains small.}
Therefore, advances in electromagnetic observation for SMBHBs, particularly the photometric variability method, could offer new opportunities for constraining scalar bosons with $\alpha\approx 0.1-0.2$, where most of the cloud remains undepleted.


\subsection{{Gravitational wave observations}}

A more promising way to identify SMBHBs is through GW observations. The presence of a boson cloud can leave an imprint on the GW waveform during either the resonant transition or ionization stages~\cite{Brito:2015oca}. 
Here, we focus on the direct detection of cloud ionization effects through GWs, taking into account the cloud depletion effects. For a single BH binary, the GW strain under the quasi-circular orbit approximation, i.e. $\dot{\Omega} \ll \Omega^2$, is given by $\tilde{h}(t)=A\cos(2\pi f t+\phi_0)$, where $f=\Omega/\pi$ is the GW frequency. Since we focus on SMBHBs at low redshifts, we also ignore the differences in the GW frequency and BH mass between the source and observer frames. Moreover, for demonstration purposes, we also ignore the polarization dependence, which is determined by the inclination angle between the binary orbit plane and the line of sight. The amplitude is then given by
\begin{eqnarray}
A=\frac{4}{d_L}(G\mathcal{M})^{5/3}(\pi f)^{2/3}\approx \frac{4 GM}{d_L} q\bar{\Omega}^{2/3}\,,
\end{eqnarray}
where {\crd $\mathcal{M}=(q(1-q) M^2)^{3/5}/M^{1/5}\approx q^{3/5} M$}
represents the chirp mass, and $d_L$ is the luminosity distance.  Since $\dot{f}\propto P_{\rm GW}+P_{\rm ion}$, ionization can have a distinct impact on the evolution of $f_{\rm GW}$ when its power dominates~\cite{Baumann:2022pkl}. As shown in Fig.~\ref{fig: ratio_left}, the condition  $\mathcal{R}\gg1$ can indeed occur across a wide range of radii at relatively smaller values of $\alpha$, provided that the GW emission or resonant transition-induced cloud depletion is insignificant.

The increased orbital decay due to ionization can significantly affect the signal-to-noise ratio (SNR) for individual sources as well. With the stationary phase approximation, the Fourier transform of the GW strain is expressed as $|h(f)|\approx A/(\dot{f})^{1/2}$. Taking into account the impact of the finite observational time $T_{\rm obs}$, the characteristic strain is given by
\begin{equation}\label{eq:hcNT}
    h_c(f)\approx A\sqrt{\min\{\mathcal{N}(f),f T_{\rm obs}\}}\,,
\end{equation}
where $\mathcal{N}(f)\equiv f^2/\dot{f}$ represents the cycle number. By deriving $\dot{f}$ using the Keplerian relation and $\text{d}x/\text{d}\bar{t}$ from Eq.~(\ref{eq:xtMct}), we obtain 
\begin{eqnarray}\label{eq:NfGW}
 \mathcal{N}(f)=\frac{f^2}{\dot{f}}=\frac{5}{96\pi^{8/3}}q^{-1}\frac{\bar{f}^{-5/3}}{1+\mathcal{R}(\bar{f})}\,,
\end{eqnarray}
where $\bar{f}=GM f$.
At the early inspiral stage, with relatively small $\dot{f}$, we have $\mathcal{N}(f)\gg f T_{\rm obs}$. The binary then acts as a continuous source, emitting GWs at a single frequency bin within the observation period, and $h_c(f)\propto f^{7/6}$. As the orbital decay accelerates significantly at the late inspiral stage, when $\mathcal{N}(f)\ll f T_{\rm obs}$, the binary behaves as an inspiraling source with evolving frequency content during the observation period. When GW emission dominates the evolution, i.e. $\mathcal{R}\ll1$, we have $\mathcal{N}\propto f^{-5/3}$ from Eq.~(\ref{eq:NfGW}), leading to  $h_c(f)\propto f^{-1/6}$. The signal becomes the strongest at the transition between these two scaling behaviors, which occurs at a critical frequency $f_{c,0}$, where $\mathcal{N}(f_{c,0})\approx f_{c,0} T_{\rm obs}$. From Eq.~(\ref{eq:NfGW}), we find $\bar{f}_{c,0}=GMf_{c,0}\propto q^{-3/8}\bar{T}_{\rm obs}^{-3/8}$, where $\bar{T}_{\rm obs}=T_{\rm obs}/(GM)$. Namely, $\bar{f}_{c,0}$ shifts to a higher value for smaller $q$ and $\bar{T}_{\rm obs}$.

\begin{figure}
    \centering
    \includegraphics[height=5.2cm]{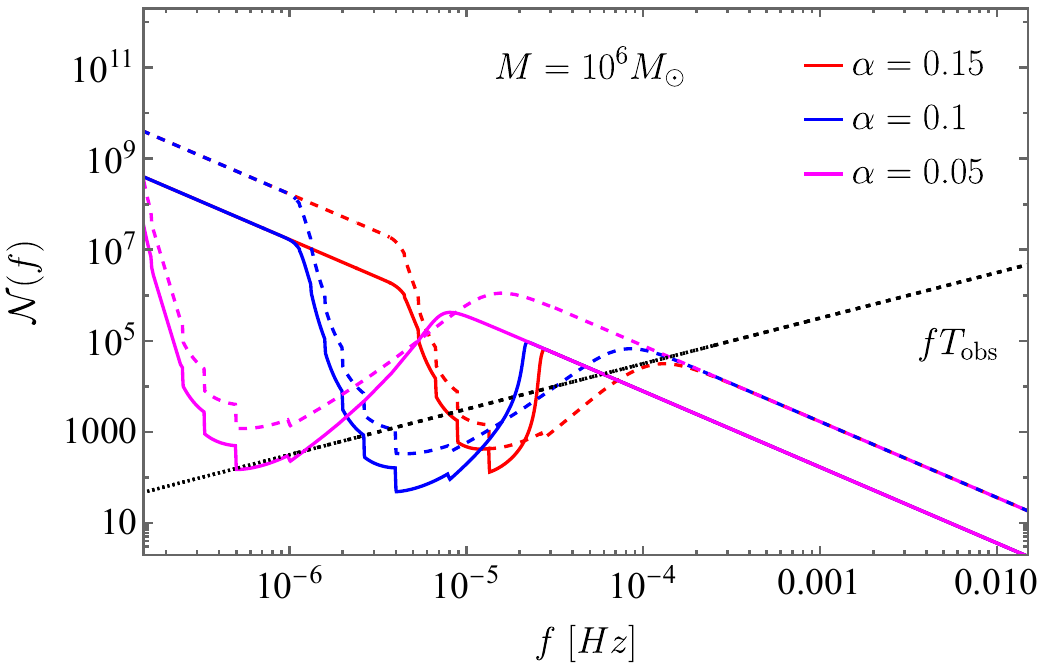}\quad
    \includegraphics[height=5.2cm]{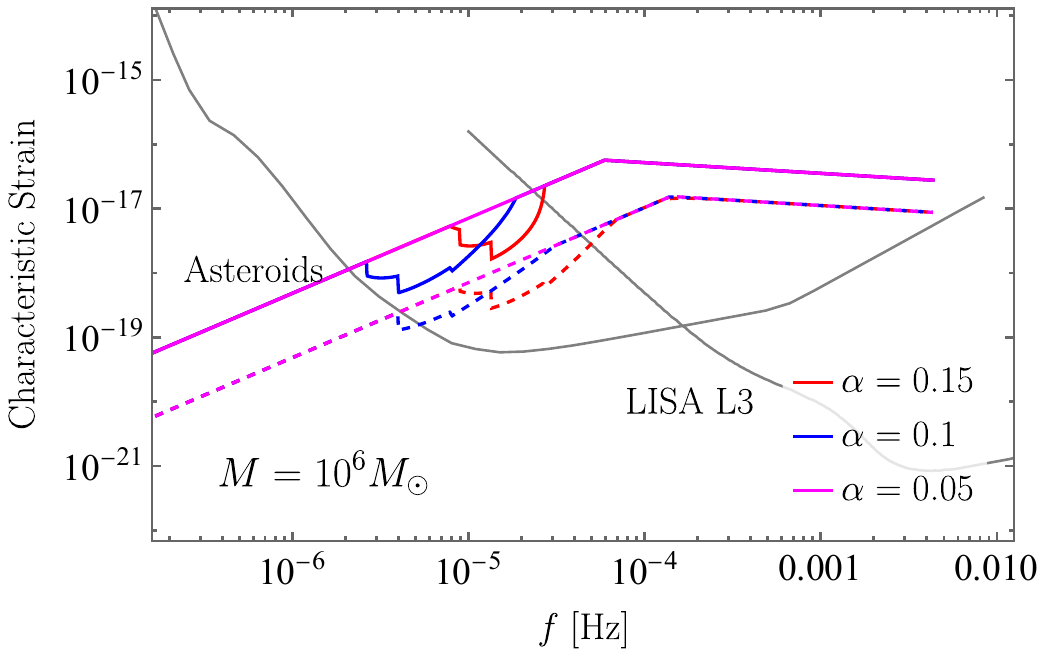}\\
    \includegraphics[height=5.2cm]{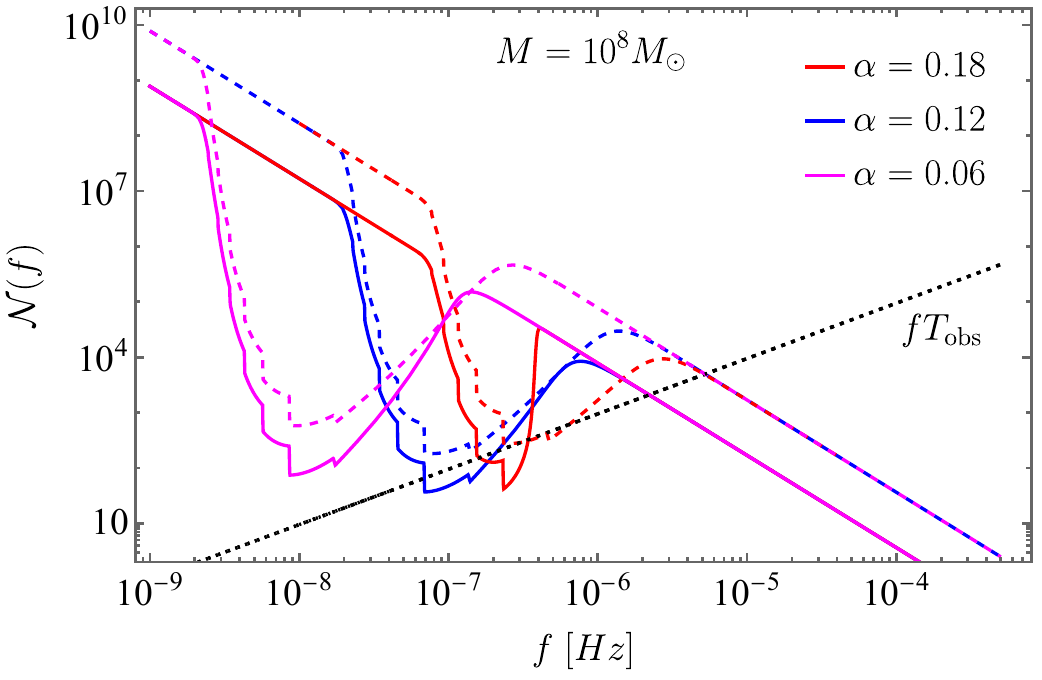}\quad
    \includegraphics[height=5.2cm]{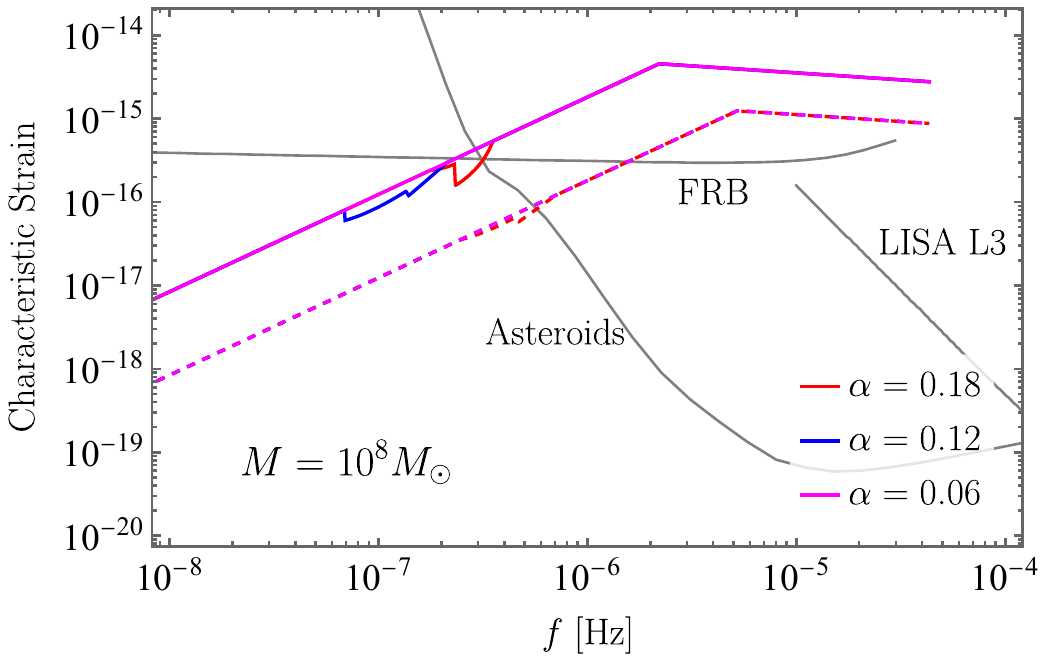} \\
    \includegraphics[height=5.2cm]{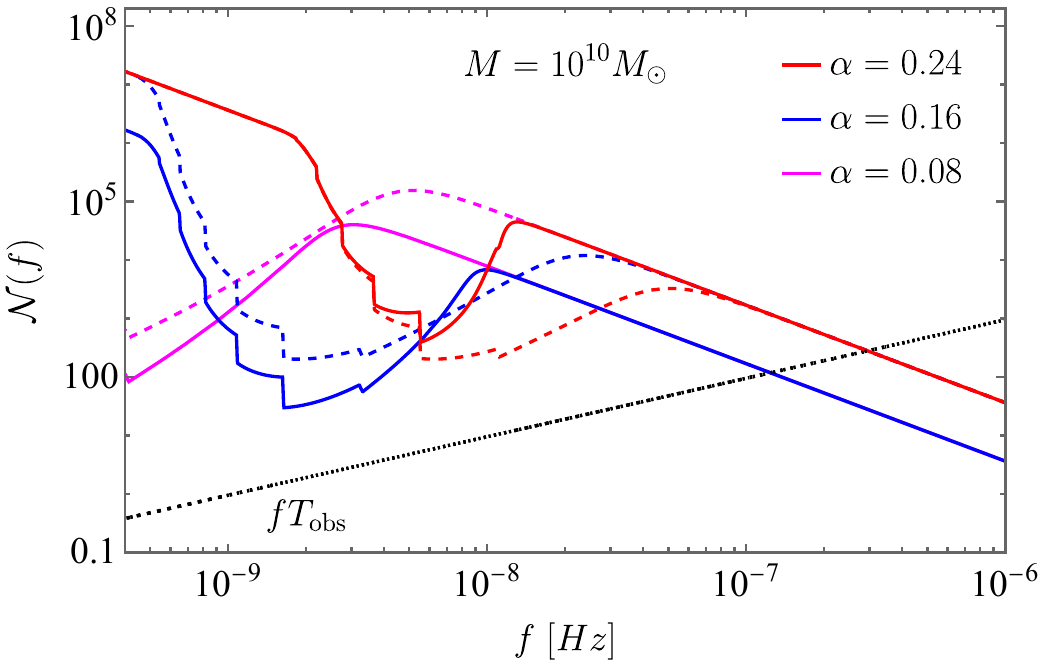}\quad
    \includegraphics[height=5.2cm]{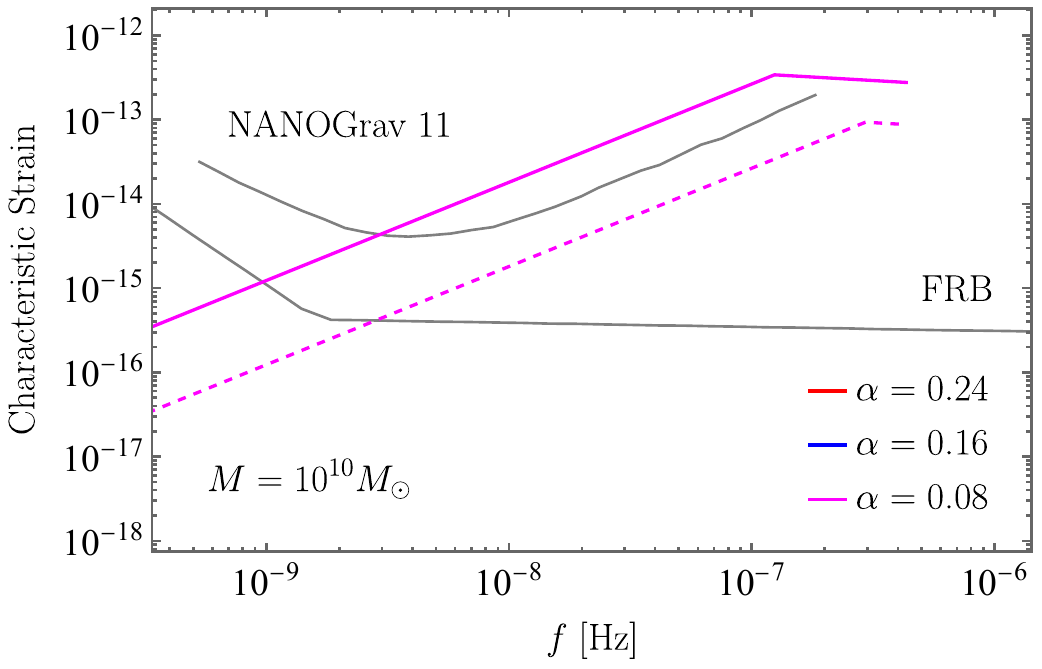}\\
    \caption{{ GW cycle number $\mathcal{N}(f)$ (left) and characteristic strain $h_c$ (right) for a single SMBHB during the inspiral stage, accounting for ionization effects. Top, middle, and bottom panels correspond to BH masses $M=10^6M_{\odot}$, $M=10^{8}M_{\odot}$ and $M=10^{10}M_{\odot}$, respectively, with the total observation time $T_{\rm obs}=10$\,yr.} The inspiral stage ends at the inner-most stable circular orbit, i.e. $\bar{r}=6$ for a Schwarzschild BH. Different colored lines represent various benchmark values of $\alpha$. Solid and dashed lines denote  binaries with $q=0.1$ and $q=0.01$, respectively. { Black dotted lines (left) show the product $f T_{\rm obs}$. Gray lines (right) indicate sensitivity curves for various experiments, including LISA L3 design sensitivity~\cite{LISA:2017pwj}, 
    projections for $\mu$Hz GW detection with asteroid test-mass~\cite{Fedderke:2021kuy} and with fast radio burst (FRB) timing assuming 1000 FRB events detected per year~\cite{Lu:2024yuo}, and the NANOGrav (11\,yr) results~\cite{NANOGrav:2017wvv}.
    All} sources are shown at a redshift of $z=0.1$, or equivalently, a luminosity distance of $d_L\approx 10^{22}\,\mathrm{km}$. 
    }
    \label{fig: GWevent_1}
\end{figure}

{ Figure~\ref{fig: GWevent_1} illustrates the cycle number $\mathcal{N}(f)$ and characteristic strain $h_c(f)$, including ionization effects, using the same benchmark values of SMBHB's $M$, $q$ and $\alpha$ as in Fig.~\ref{fig: ratio_left} in the relevant frequency bands. When ionization power begins to dominate over GW emission during the early inspiral stage, i.e. $\mathcal{R}\gtrsim 1$,   $\mathcal{N}$ decreases significantly due to an enhanced orbital decay rate. If this reduction is substantial enough that $\mathcal{N}$ falls below $f T_{\rm obs}$, the nearly monochromatic signal is changed earlier in the evolution. The critical frequency for $\mathcal{N}$ and $f T_{\rm obs}$ crossing, denoted as $f_{c,1}$, then shifts to a lower value.}
As the frequency continues to increase, the signal strength becomes strongly suppressed and decays more rapidly than $f^{-1/6}$, due to the more rapid increase in ionization power. The characteristic strain reverts to the GR prediction at $f_{c,2}$ in higher frequency (or smaller $x$) when $\mathcal{N}>f T_{\rm obs}$ or $\mathcal{R}<1$ due to the reduced ionization power.
By substituting the value of $\mathcal{N}$ from Eq.~(\ref{eq:NfGW}), the first critical frequency $f_{c,1}$ can be determined by $\mathcal{R}(x(\bar{f}_{c,1}))=(\bar{f}_{c,0}/\bar{f}_{c,1})^{8/3}$, where $\bar{f}_{c,0}$ corresponds to the critical frequency at the same $M$ and $q$. 

{For the $M=10^6 M_\odot$ case, the critical frequency $f_{c,1}$} occurs around $10^{-5}\,$Hz, which is below the most sensitive region of LISA but could be well probed by $\mu$Hz GW detections, such as proposals with asteroid test-masses. Since the ratio $f_{c,0}/f_{c,1}$ is determined by $\mathcal{R}$, the values of $f_{c,1}$ are mostly sensitive to $\alpha$, shifting to a lower value for smaller $\alpha$ due to the $\alpha^{-5}$ enhanced $\mathcal{R}$. 
{ However, for the smallest $\alpha$ value case, corrections to the GR prediction become nearly negligible because $\mathcal{R}$ remains insufficiently large at small $x$ to reduce $\mathcal{N}$ below $f T_{\rm obs}$.}
The position of $f_{c,2}$ is highly sensitive to the mass ratio. For the case of $q=0.1$, cloud depletion during ionization is much more efficient compared to the $q=0.01$ case, as shown in Fig.~\ref{fig: ratio_left}. Consequently, ionization-induced corrections span a narrower range of GW frequencies, with a smaller value of $f_{c,2}$. In contrast, for the lower mass ratio case, although the overall signal is suppressed by $q$, ionization could affect a slightly broader frequency range.

{ For increasing BH mass, the value of $\bar{T}_{\rm obs}$ decreases for a fixed $T_{\rm obs}$. This then requires a larger $\mathcal{R}$ to sufficiently reduce  $\mathcal{N}$, and $\bar{f}_{c,1}$ also shifts to a higher value.}
Additionally, the allowed range for $\alpha$ shifts to larger values, { limiting the potential enhancement of the ionization-to-GW ratio $\mathcal{R}$.} These factors make it more challenging to observe the corrections from ionization. 
{ For $M=10^8M_{\odot}$, cloud effects generally weaken. In particular, for small $q$ and $\alpha$ values, the predictions may fully revert to the GR case, since $\mathcal{N}$ never fall below $f T_{\rm obs}$ in the relevant frequency band. 
For $M=10^{10}M_{\odot}$, the corrections are invisible for all benchmark values of $\alpha$ and $q$ due to the large values of $\mathcal{N}$, resulting in overlapping $h_c$ curves for different values of $\alpha$. The only possibility of observing such effects would require significantly increasing $T_{\rm obs}$ beyond the current order of 10\,yr.}

Overall, beyond the changes in GW frequency noted in Ref.~\cite{Baumann:2022pkl}, 
ionization effects could also substantially reduce the SNR of individual SMBHB sources, rendering them undetectable in the originally {\crd detectable} observational band. This is especially true for relatively smaller BH mass $M$, complementing electromagnetic observations. 

Finally, let's briefly discuss the potential impact of cloud ionization on  stochastic GW backgrounds (SGWBs) formed by a population of SMBHBs. The characteristic strain of SGWBs can be derived by integrating the merger number density of the SMBHBs with the spectral energy density  $\mathrm{d}E_{\rm GW}/\mathrm{d}f$ of the binaries. If the SGWB arises from SMBHBs, the observed spectrum is dominated by comparable mass binaries with $q\sim 1$~\cite{Chen:2020qlp, NANOGrav:2023hfp, NANOGrav:2024nmo}, and thus the spectrum's shape primarily depends on the frequency dependence of $\mathrm{d}E_{\rm GW}/\mathrm{d}f$ for these binaries. In cases where $q\sim 1$, the boson cloud would experience more complex evolutionary processes, such as cloud transfer between the two BHs~\cite{Wong:2020qom,Ikeda:2020xvt,Guo:2023lbv,Guo:2024iye}, rather than the simple bound state transitions discussed here. Therefore, the effects we focus on, which relates to the subset of binaries with $q\lesssim 0.1$, provide only a minor contribution to the overall SGWB spectrum.

\section{Summary}
\label{sec:summary}


In this work, we explore the evolution of the cloud-binary system for SMBHBs by accounting for their potential interactions with the environment and examining the resulting observational consequences.

In Sec.~\ref{sec:resonance}, we investigate the efficiency of resonant transition to bound states by considering the astrophysical evolution history of SMBHBs. 
If the transition occurs at a large separation before the system enters the GW radiation stage, corresponding to $\alpha\lesssim \alpha_{\rm gr}$ given in Tab.~\ref{tab:alpha}, the resonant transition can be significantly influenced, due to contributions from the additional energy loss channels to the orbital evolution rate $\mathcal{G}$ in Eq.~(\ref{eq:rateG}). Specifically, we employ Eq.~(\ref{eq:Gtpeak}) for a conservative estimate of $\mathcal{G}$, utilizing statistical distributions of the peak evolution time for SMBHBs in Ref.~\cite{Chen:2020qlp}. As illustrated in Fig.~\ref{orbit_evolution}, for $\alpha\lesssim \alpha_{\rm gr}$, the parameter space for non-adiabatic transitions significantly expands. Even when the adiabatic condition is satisfied, the {\crd initial state} population resulting from resonance-breaking is also greatly increased. This finding suggests that the constraint on the orbital inclination angle -- necessary for maintaining a sufficient cloud for direct detection at a later stage -- is notably relaxed compared to the limits found in Ref.~\cite{Tomaselli:2024bdd}. This highlights the critical importance of considering realistic binary environments in the detection of boson clouds, complementing the nonlinear effects on cloud depletion caused by geometric changes from cloud backreaction~\cite{Takahashi:2021yhy}.  
In Sec.~\ref{sec:ionization}, we examine ionization effects, noting that both ionization power in Eq.~(\ref{eq: P_ion}) and dynamical friction power in Eq.~(\ref{eq: P_DF}) share similar scaling behaviors. Fig.~\ref{fig: FF1} shows that although these power are alike near the peak region, ionization power decreases much more rapidly at larger radii, highlighting a distinct difference from dynamical friction.
We also explore cloud depletion during the ionization stage in detail. The cloud depletion exhibits different behaviors and dependencies on $q$ and $\alpha$, depending on whether ionization power or GW power dominates. Fig.~\ref{fig: ratio_left} indicates that cloud depletion does not significantly reduce the ionization-to-GW power ratio only when $q$ is small and $\alpha$ is intermediate.  

We then examine the potential imprints of boson clouds on SMBHB observations in Sec.~\ref{sec:observation}. We focus on the direct detection of cloud ionization effects, considering the previously discussed cloud depletion effects.
For electromagnetic observations of SMBHBs, we consider the orbital period decay rate as an observable. When ionization power dominates, the decay rate is described by Eq.~(\ref{eq:Tdot}). As illustrated in Fig. \ref{fig:T_dot}, for  SMBHBs with $M\gtrsim 10^8 M_{\odot}$, the maximum decay rates {occurring at $\alpha\approx 0.1-0.2$ do fall within observationally favored parameter ranges for the photometric variability method--making this approach} 
particularly promising for probing such subtle changes in the future.
For GW observation of a single binary, in addition to the distinct impact of ionization on the evolution of the GW frequency, the characteristic strain of the GW can be significantly affected.  Specifically, substantial ionization power may shift the transition from a continuous source to an inspiraling source to a lower critical frequency. 
This can substantially reduce the signal-to-noise ratio of individual sources, rendering them undetectable in the originally {\crd detectable} observational band.
Fig.~\ref{fig: GWevent_1} illustrates that the ionization effects on the waveform, for various benchmark values of { $\alpha\approx 0.1-0.2$}, may be detectable around the $\mu$Hz band for $M\lesssim 10^8 M_{\odot}$, complementing electromagnetic observations.

Our initial study of environmental effects on resonant transition {\crd relies} on a rough estimate to illustrate an expanded parameter space that allows sufficient cloud retention for direct detection. For a closer connection to observations, a more in-depth exploration of the interplay between the the statistical distributions of SMBHBs, environmental effects, and $\alpha$ is needed. In the cases where resonant conversion is rendered inefficient due to environmental effects, the potential influence of off-resonant transitions from the {\crd initial state} to other bound states~\cite{Tong:2022bbl} deserves further investigation.
Moreover, it would be interesting to understand the evolution of boson clouds around comparable mass SMBHBs in the presence of additional energy loss channels. A more comprehensive investigation along this line could provide insights into searching for boson clouds through observations of SGWBs~\cite{Guo:2025pea}. This remains an area for future research.

\vspace{0.1cm}
\section*{Acknowledgements} 
\vspace{-0.1cm}

We would like to thank Yun-Feng Chen, Ao-Yun He, Yan-Rong Li, and Guan-Wen Yuan for valuable discussions on astrophysical evolutionary histories and observations of SMBHBs.
J.R. is supported in part by the National Natural Science Foundation of China under Grant No. 12435005.


\clearpage
\newpage

\bibliographystyle{JHEP}
\bibliography{reference}

\end{document}